\documentclass[aps,prd,twocolumn,groupedaddress,nofootinbib,showpacs]%
{revtex4-1}

\usepackage{amsfonts}
\usepackage{graphicx}
\usepackage{bm}
\usepackage{amssymb}
\usepackage{amsmath}
\usepackage[T1]{fontenc}
\newcommand{\scrif}{{{\mathcal I}^{+}}}

\newcommand{\C}{{\mathbb C}}
\newcommand{\R}{{\mathbb R}}

\newcommand{\Q}{{\mathcal Q}}
\newcommand{\RR}{{\mathcal R}}

\newcommand\tn{{\tilde n}}
\newcommand\tm{{\tilde m}}
\newcommand\tom{{\tilde\omicron}}
\newcommand\tomicron{{\tilde\omicron}}
\newcommand\tio{{\tilde\iota}}
\newcommand\tiota{{\tilde\iota}}
\newcommand\tnab{{\tilde\nabla}}
\newcommand\txi{{\tilde\xi}}
\newcommand\teth{{\tilde\eth}}
\newcommand\tthorn{{\tilde{\thorn}}}

\newcommand\sign{{\mathop{\rm sgn}}}

\def\const{{\rm constant}}

\def\omicron{{o}}
\def\thorn{\hbox{\TH}}
\newcommand{\rmi}{{i}}
\newcommand{\rmd}{{d}}

\newcommand{\sech}{\mathop{\rm sech}\nolimits}
\newcommand{\zz}{{\mathfrak z}}

\begin{document}

\title{Energy--momentum and asymptotic geometry}

\author{Adam D. Helfer}
\email[]{helfera@missouri.edu}
\affiliation{Department of Mathematics and Department of Physics \& Astronomy,
University of Missouri,
Columbia, MO 65211, U.S.A.}

\date{\today}

\begin{abstract}
I show that radiative space--times are not asymptotically flat; 
rather, the radiation field gives rise to holonomy at null infinity.
(This was noted earlier, by Bramson.)  
This means that, when gravitational radiation is present, {\em 
asymptotically covariantly constant} vector fields do not exist.  On 
the other hand, according to the Bondi--Sachs construction, a weaker 
class of {\em asymptotically constant} vectors does exist.  
Reconciling these concepts leads to a measure of the scattering of
matter by gravitational waves, that is, bulk exchanges of energy--momentum between the waves and matter.  Because these bulk effects are potentially larger than the tidal ones which have usually been studied, they may affect the waves' propagation more significantly, and the question of matter's transparency to gravitational radiation should be revisited.
While in many cases there is reason to think the waves will be only slightly affected, some situations are identified in which the energy--momentum exchanges can be substantial enough that a closer investigation should be made.  In particular, the work here suggests that gravitational waves produced when relativistic jets are formed might be substantially affected by passing through an inhomogeneous medium.
\end{abstract}

\pacs{04.20.Ha
04.30.Nk
95.30.Sf
}

\maketitle

\section{Introduction}

The aim of this paper is to present limited but clean results on the interaction of gravitational waves with matter.  I shall explain how waves exchange energy--momentum with  small amounts of matter in the waves' radiation zone. 
(Throughout this paper, ``matter'' means anything with stress--energy; in particular, it includes electromagnetic radiation.
I consider only outgoing radiation; of course, one could time-reverse the treatment to obtain results for incoming waves.)

Here the {\em radiation zone} of an isolated source will be a (usually finite, large) regime around it in which 
certain elements of
the space--time geometry can be well-approximated by the leading terms in the Bondi--Sachs asymptotic expansions \cite{BVM,Sachs1962a}.  
In particular the system need not be ideally isolated, that is, the Bondi--Sachs geometry need not approximate the physical geometry indefinitely far out.  When the waves leave this zone (because they begin to encounter curvature from other sources), other physics takes over.

Of course, it is well-known that waves' tidal effects can alter the energy--momentum of one small body relative to another nearby; here, however, the project is to understand how each body's local energy--momentum contributes to the system's total.  Thus we are interested in {\em bulk} energy--momentum exchanges between matter and radiation; tidal effects will be the differentials of these.\footnote{A word about the analogy with electromagnetism is in order.  When we think of electromagnetic waves encountering matter, we usually have in mind matter which is (macroscopically) neutral.  Then the main macroscopic effects come from its polarizability, which is analogous to a tidal distortion in gravity.  But the parallel with macroscopically neutral matter is not the correct one, since
mass, the relativistic analog of charge, comes with only one sign.  It would be better to think of electromagnetic waves encountering distributions of charge.}

The 
premise will be that
the Bondi--Sachs formalism gives a convincing treatment of the total energy--momentum $P_a^{\rm Bondi-Sachs}$ of systems which are idealized as perfectly isolated (that idealization being reflected in that $P_a^{\rm Bondi-Sachs}$ is defined  strictly at null infinity).  The aim is to extend this construction inwards, to finite
points in the radiation zone.  In this sense the work here is a step towards treating energy--momentum quasilocally.

Another way of viewing this is as a search for a (limited) general-relativistic analog of potential energy --- we 
seek a way of relating the  
energy--momenta of localized objects in the radiation zone (which take values in the cotangent bundle) to the energy--momentum of the entire system (which exists in a sort of ``cotangent space at infinity'').

We do not 
know enough to solve such problems from first principles, or even to be confident that they have solutions.\footnote{The
twistorial approach suggests that the quasilocal kinematic quantities will not generically be energy--momentum and relativistic angular momentum, derived from the Poincar\'e group, but quantities modeled on de Sitter or anti-de Sitter symmetries.}  
Here the idea is not to guess at an overarching formalism, but to use the special properties of the radiation zone to guide us to a physically plausible approximate treatment. 

We will see that this zone has a distinctive geometry which codes at once the gravitational radiation and the difficulties in relating local to global measures of energy--momentum.  Examining the Bondi--Sachs construction with this understanding will give a way to resolve those difficulties, and so to treat energy--momentum in the radiation zone.

This approximate treatment is useful:  it clarifies conceptual points about the physical interpretation of the radiation zone
and gives computations of scattering.  
The core of the physics is the nonlocality of the total, general-relativistic, energy--momentum.  
This is on one hand closely connected with a(n unsubtle) scaling property implied by Sachs peeling, and one the other with the (arguably subtle) fine gauge control provided by the Bondi--Sachs approach. 

I began by characterizing the present results as limited; this is because the treatment requires the matter perturb the radiation-zone geometry only slightly.  This will mean that, while interesting effects will be uncovered, they will generally be fractionally small where the treatment is valid.  The question of whether they can be more substantial when the hypotheses here are relaxed
will have to be answered by other means.  What the approach here does do is to identify potentially interesting cases of energy--momentum exchange.

\subsubsection*{Main results}

The main new formulas describe how local measures of energy--momentum in the radiation zone are related to the total energy--momentum.  For example, the contribution of a test particle of mass $\mu$ freely falling along a geodesic $\gamma$ to the total energy--momentum will change as the particle encounters radiation, the rate of change along the trajectory being denoted
\begin{eqnarray}\label{fallscat}
  \dot\gamma ^bD_b(\mu\dot\gamma _a)&=&\mu\left(\kappa ' l_b-\sigma 'm_b\right)\left( l_am^c-
m_al^c\right)
\dot\gamma ^b\dot\gamma _c\nonumber\\
&&\quad +\,\text{conjugate}\, . 
\end{eqnarray}
where the spin-coefficents $\sigma '$, $\kappa '$ code the radiation, and the Bondi--Sachs tetrad vectors $l^a$, $m^a$ its outward-propagating transverse character.   
Integrating (in a suitable sense) this will give us a measure of the scattering the particle suffers owing to radiation.  

The Bondi coordinates are $(u,r,\theta ,\phi )$, where $u$ is the {\em Bondi retarded time} (one can think of the $u=\const$ hypersurfaces as the outgoing wave-fronts; they are null and $l_a=\nabla _a u$), the coordinate $r$ is an affine parameter along the null geodesics generating those fronts, and $\theta$, $\phi$ are angles.  
In the asymptotic regime, one has $\sigma '\sim N/r$, $\kappa '\sim \teth N /r$, where 
$\teth$ is a certain angular derivative, and $N=N(u,\theta ,\phi)$, the {\em Bondi news}, 
is essentially a potential for the radiative components of the curvature.  Thus contributions to scattering from high-frequency wave-packets in the news will tend to average out.  More precisely, this suppression will occur if neither the value of $r$, nor the dilation ${\dot\gamma}^a\nabla _au$ of the Bondi retarded time relative to the particle's proper time, nor the angular dependence of the news, is significant on the portion of the particle's world-line extending over 
a period of oscillation.  
To avoid such cancellations, 
the particle must pass through an angle on the sphere of directions outwards from the source over which significant contributions from the news can accumulate.
This means
either sources with very strong angular dependences, or particles moving rapidly enough past them that they subtend a significant angle over a period of oscillation.
Note that for low-enough frequency components, and in particular sources with ``memory,'' the cancellation mechanism does not apply.

More generally, for any distribution of matter in the radiation zone, we find that the rate of conversion of material contributions to the total energy--momentum to gravitational-wave contributions, per unit time per unit volume, is
\begin{equation}\label{seform}
\frac{d{\mathcal P}_d}{d\tau}=  T_{ab}(\sigma ' m^a-\kappa 'l^a)(l_dm^b-m_dl^b)
    +\mbox{conjugate}
\end{equation}
where $T_{ab}$ is the stress--energy.  (It should be emphasized that this does not mean that matter is created or destroyed; what is changing is the matter's contribution to the total energy--momentum of the system, just as a mass in a Newtonian potential contributes differently to the total energy, depending on its position.)  That this depends only on the stress--energy and not on other characteristics of the matter can be viewed as a compatibility of the approach here with the weak equivalence principle.\footnote{Contrast this with the usual view of tidal effects, where for instance the local energy-exchange between two masses on a spring and gravitational waves depends on the stiffness of the spring, not just its mass.  To reconcile these views, note 
that the relative difference in energy between two superficially similar springs of different stiffnesses is only a tiny fraction of the springs' rest-energies.  Here, in considering bulk effects, it is the springs' relativistic energy--momenta which we track, and we see how small a fraction of this tidal effects are.}

For general distributions of matter, there are more possibilities for energy--momentum exchange than for freely-falling test masses.  Most importantly, the stress--energy tensor of the matter may have significant local time-dependence and there is the possibility of resonant beating against the gravitational waves to drive intervals of secular exchange.  (The possibility of electromagnetic and gravitational waves beating against each other was suggested long ago, by Gertsenshtein~\cite{Gertsenshtein1962}.  However, the set-up here is much more general and the effects here are  different from his.)  

Closely related to this is the question of what sorts of redirection and absorption of gravitational-wave energy--momentum are possible by matter.
(Note that this is not quite the same as redirecting or absorbing the waves; also because bulk rather than tidal effects are concerned it differs from earlier investigations, for example that of Press \cite{Press1979}.)  This issue is complicated by the nonlocality of the energy--momentum, but we do find that there are some circumstances in which nearly local statements are possible and matter can alter the energy--momentum by  terms proportional to the waves' outgoing null covector $l_a$.

Another possibility for effects which are not suppressed by averaging occurs with ``memory,'' in this case a net difference in Bondi shear between two non-radiating regimes.  Substantial differences in shear are expected in particular for relativistic jets.  We find that such radiation passing through non-relativistic matter may lose momentum, perhaps enough to significantly degrade the waves.  This could affect their detectability; compare ref. \cite{BP2013}.

\subsubsection*{Implications for propagation}

Verifying that energy--momentum can be exchanged between matter and radiation is gratifying but unsurprising.  It does, however, raise important questions about propagation:  for these exchanges should cause back-reactions on the waves, and this calls us to reexamine the
common claim that passage through matter does not alter the waves (except for background-curvature effects, or in extraordinary circumstances).  
Ultimately, detailed gravitational-wave astronomy, measuring many parameters of sources, will require better than per-cent-accuracy knowledge of certain features of the wave-forms \cite{LOB2008}; for this, even relatively small effects need to be seriously considered.

The usual arguments that matter is transparent to gravitational waves depend in part on estimates about how energy can be exchanged, but those estimates have been based
on an implicit assumption that one need only consider tidal effects \cite{Thorne1982,Thorne1987}.  As those are only the differentials of whatever bulk effects are present, the question of transparency needs to be reconsidered.

Unfortunately, the techniques here track energy--momenta, and not wave-forms, so they do not give direct information about propagation.  
However, because the emitted energy--momentum is quadratic in the radiation,
it is reasonable to suppose that the orders of magnitude of the fractional changes in wave-form and the waves' energy--momentum, due to intervening matter, are the same.
Now, because the basic assumption here is that the geometry of the radiation zone is well-approximated by the Bondi--Sachs asymptotics, which treat vacuum space--times (or, at most, those with an electromagnetic radiation field), any back-reaction effects should be small.
But there is no general reason to think that these effects are limited in principle to be well below the per-cent level.

As pointed out earlier, in many cases the waves will oscillate much more rapidly than the stress--energy changes, and for these the net energy--momentum exchange will tend to be suppressed.  While we do not understand just how this averaging should affect the waves, it does seem plausible that in these cases there will be a significant suppression of back-reaction effects.  
Recall, though, that in some  important cases (as with jets) there may be effects which are not suppressed by averaging.

The question of just which degrees of freedom of the signals could be affected  by back-reaction is critical; it could well be that in many cases the effects of encounters with matter could easily be separated.  This would open the possibility of extracting {\em more} information from gravitational waves.
We need careful analyses of propagation to investigate and clarify this matter.

I have so far described what will be done; I now sketch how it will be done.

\subsection{The main ideas}

The main ideas of this paper turn on formalizing the idea of a radiation zone and on its nonlocal geometry.  The approach of Bondi and Sachs is used very strongly.  While this certainly overlaps with the less formal notion of a radiation zone based on treating waves as perturbations of a background, the fine gauge control of the Bondi--Sachs approach is essential.

\subsubsection*{The radiation zone and radiation-dominance}

It is probably fair to say that Bondi's approach was aimed at finding a suitable, fully covariant, characterization of the radiation zone of a system idealized as perfectly isolated.  He was led, by previous investigations, to hypothesize that in this regime the geometry should admit a certain asymptotic coordinate system
$(u,r,\theta ,\phi )$ (with $u$ a ``retarded time,'' the $u=\const$ hypersurfaces being null and opening outwards, the ``radial'' coordinate $r$ being an affine parameter up the null generators of these hypersurfaces,\footnote{Actually, Bondi used a luminosity distance, not an affine parameter, but affine parameters are simpler for most purposes and have been used in most subsequent work.  Also Bondi assumed axisymmetry; the generic case was treated by Sachs.}  and $(\theta ,\phi)$ angular variables on the sphere), with respect to which the metric would have a certain asymptotic expansion as $r\to\infty$.  Penrose then showed that these conditions could be recast as the existence of a conformal boundary; this led to an elegant formalism and also shifted the focus of work from the radiation zone to null infinity itself, for much could be said about the limiting forms of quantities there.

Our view 
will be closer to Bondi's original one, however.  We shall say a physical system {\em admits a radiation zone } if 
it has a region 
in which the geometry is well-approximated by the leading terms in the Bondi asymptotic expansion.  (We need no physical hypotheses on the rest of the system or the Universe, although we may, as a mathematical convenience, imagine embedding the zone in an auxiliary space--time extending to null infinity.)  
This parallels the notion of a radiation zone in special-relativistic electromagnetism, as a regime in which the field is well-approximated by its radiative term.

While this definition is 
natural, it has an important novel feature.  It does not restrict the radiation field to be weak, it {\em does} mean that any matter present should perturb the geometry of the radiation zone only slightly.  This is the most important restriction on the approach here.  Notice that it is
opposite
in spirit to the usual assumption that the radiation field only perturbs the matter infinitesimally.

We may call this the {\em radiation-dominated} regime, and contrast it with the more usual {\em matter-dominated} one.  It would evidently be a natural limit to consider, if only for conceptual reasons.  
One might wonder, in fact, if  it is only of academic interest, since 
gravitational waves are generally very weak --- but the coupling of matter to curvature is also weak, being mediated by the gravitational constant $G$.  So just when does radiation-dominance apply?

To answer this, we must 
specify which elements of the physical regime's geometry must be well-modeled by the Bondi--Sachs asymptotics.  
 
However, because we do not have a fundamental understanding of quasi-local kinematics, we have no way of knowing all the geometric structures which might turn out to be relevant.
What we can do is point to the minimal set of elements of the geometry which are involved in our construction.    
We shall find that there are plausible situations in which radiation-dominance, in this sense, holds.

Finally, a comment on invariance is in order.
Let us start with a parallel case, an electromagnetic radiation zone in special relativity.  In such a zone, the field appears to good approximation to be radiative --- that is, to be outgoing transverse waves.  This zone will certainly {\em not} be Poincar\'e-invariant.  For one thing, translations will move one out of the zone.  But also, even locally, large enough boosts will destroy the approximation that the waves are transverse --- non-transverse terms, which are small in the frame of the radiation zone, will not be small in other frames.  (The zone does have an approximate invariance, for ``small enough'' translations or Lorentz motions.)  The usefulness of a radiation zone is not that it is invariant, but just the opposite, that it gives a distinguished frame rendering the field simple.

Parallel comments apply to the gravitational case, where the relevant asymptotic symmetries form the Bondi--Metzner--Sachs (BMS) group.  Any structure which is universal to isolated radiating space--times will be BMS-invariant, but specific radiation zones will not be.  
The construction here relies in particular on identifying the outgoing direction in which the waves propagate.  This is subject to an ambiguity, which is the same as the approximate symmetry group, and varies inversely with the extent of the radiation zone.  (Note that in practical cases this ambiguity is of the order of the angle subtended by the source as viewed from the inner points in the zone.)

\subsubsection*{Nonlocality and scattering}

The core issues can be brought out by considering a 
prototypical problem:
Suppose a test particle falls freely through the radiation zone.  Its energy--momentum then remains, in its own frame, unchanged.  So how, or in what sense, can one say it has been scattered?

Historically (up to around 1957 \cite{Kennefick2007}), concerns like this were used to argue that gravitational radiation had no physical significance.  The response was to look to tidal effects of the waves, as in-principle (and now, we hope, in-practice) local observable properties.  
That did help convince workers of the waves' physical reality, which was a critical advance.  But from the point of view of the scattering problem, it represented a retreat, for tidal effects are only the {\em differentials} of whatever {\em bulk} scattering is present.
It is the bulk effects which we want to analyze.

The local energy--momentum of a test particle (or other localized material body) at an event $q$ in the radiation zone takes its value in the cotangent space $T_q^*$, so if we wish to compare the energy--momenta of such particles, we need a way of identifying the cotangent spaces --- in mathematical terms, a parallelism of the cotangent bundle of the radiation zone.

One's first thought might be to try to use parallel propagation (over paths restricted to the zone) to define a parallelism, at least in a limiting sense for distant enough events.  However, this fails.  Sachs peeling implies that {\em precisely when gravitational radiation is present, there are holonomic obstructions which persist even in the limit of more and more distant paths.}  An equivalent statement is that {\em when gravitational radiation is present, asymptotically covariantly constant vector and covector fields do not exist.}\footnote{Essentially this observation was made earlier, by Bramson \cite{Bramson1975}.}  In this sense, radiative space--times are {\em not} asymptotically flat.

What, then, can we do to find a parallelism?  We shall look to the 
construction
of the Bondi--Sachs energy--momentum $P_a^{\rm Bondi-Sachs}$ for a guide.
This energy--momentum takes values in a certain asymptotic covector space $T^*$.  
Roughly speaking, this space can be given as follows:\footnote{The construction is more easily done in spinor terms, and these are used in the body of the paper, but for conceptual reasons it is sketched here with covectors.}

\begin{enumerate}

\item Start with smooth covector fields $\xi _a$ on the Bondi chart.

\item Discard certain components (with respect to the Bondi chart) of the fields.

\item Impose certain differential equations, derived from but weaker than the covariant-constancy conditions, on the remaining components. 

\item Look for asymptotic solutions as 
$r\to\infty$.

\end{enumerate}

\noindent The result is the  
space $T^*$.  The most important nonlocal aspect of the construction is that the differential equations imposed are elliptic in the angular variables $(\theta ,\phi )$, so that the four-dimensional solution-space arises from requiring regularity over the sphere of directions; locally, the system of equations is underdetermined.

We will extend this construction so that it provides a well-defined four-dimensional family of covector fields in the radiation zone. 
Two changes are necessary:  the (discarded) components of $\xi_a$ are recovered from certain components of the covariant-constancy equations, and a further differential equation, propagating the fields parallel along the outgoing Bondi null congruence, is imposed.

The result of this construction is a four-dimensional space $T^*_{\rm asymptotic}$ of covector fields which we take, by definition, to be {\em asymptotically constant}, although they will {\em not } be asymptotically {\em covariantly } constant.  These define the parallelism.

It is precisely the deviation of the asymptotically constant covector fields from asymptotic covariant constancy which is responsible for, and codes, the exchange of energy--momentum between matter and the gravitational radiation field.  
For instance, a test particle's energy--momentum $P_a$ at a particular event $q$ may be identified with the asymptotically constant covector field $\xi _a$ with $\xi _a(q)=P_a$.  As the particle falls through the radiation zone, however, one will not be able to maintain this equation with a single asymptotically constant covector field $\xi _a$, for the particle's equation of motion is local, but the fields $\xi _a$ are determined nonlocally.  
Just this gives the change in the particle's contribution to the total energy--momentum, as an element of $T^*_{\rm asymptotic}$, as the particle moves.

\subsubsection*{Limitations}

I have sketched the underlying ideas of the analysis, and shown how this calls for a reexamination of gravitational-wave propagation.  This was in part because the techniques here give us no direct information about wave-forms.  However, there was also another limitation, whose force is important to understand, which is due to the approach and ultimately the lack of a general, quasilocal, kinematics.

The basic logical architecture of this approach is:

\begin{enumerate}

\item
We defined a radiation zone as a regime in which certain elements of the space--time geometry were well-modeled by their leading Bondi--Sachs expansions.  In this regime, a consistent treatment of energy--momentum 
was possible --- again, to leading order in the expansions.
The plausiblity of this treatment rests on the 
existence of
a radiation zone in the sense defined, because it is in this zone that the construction of the asymptotic covectors precisely compensates for the holonomic obstructions caused by Sachs peeling of the radiation field.

\item
While usually the Bondi--Sachs analysis has been applied to space--times which are vacuum, or have only electromagnetic stress--energy, in the asymptotic regime, here the whole point of the program is to consider what might happen for different sorts of matter in the radiation zone.  
At the same time, the main assumption is that certain elements of the geometry cannot deviate too much from those of the vacuum case.

\end{enumerate}

So by its nature, the approach here cannot describe any large relative changes in gravitational radiation or its energy--momentum due to matter.  As soon as such changes cause substantial changes in the relevant elements of the asymptotic geometry, the treatment has no clear justification.
It is not clear whether these limitations are only features of the arguments used here, or are really absolute physical restrictions. 
This applies in particular to the questions of how strongly the propagation of gravitational waves, or the energy--momentum they carry, may be affected by matter.

\subsection{Comparison with earlier work}

This paper is most naturally viewed as taking up lines of thought which were interrupted some time ago.  Most directly, these are the study of radiation zones in the form initiated by Bondi, and the problems of energy--momentum transfer which were part of the debate on the significance of gravitational waves \cite{Kennefick2007}.  
As discussed above, historical accidents turned research in different directions, focusing attention on null infinity rather than the radiation zone, and on tidal effects rather than bulk scattering.

The interaction of gravitational waves with matter has received almost no direct attention within the Bondi--Sachs framework (although the analysis of the local energetics of tidal effects was a key step in Bondi's work).  These interactions have received some consideration within linearized-perturbation frameworks.  
There are two standard reviews touching on this, by Thorne \cite{Thorne1982} and by Grishchuk and Polnarev \cite{GP1980}.  I will comment on those; for further work, see the references in those, and also in ref.~\cite{Thorne1987}.

Thorne discusses the absorption or dispersion of waves, with the assumption that those processes are due to tidal effects, and concludes that in real astrophysical situations they are totally negligible.  By contrast, Grishchuk and Polnarev discuss some processes which could, in principle, include bulk effects:  (a) an Einstein--Maxwell system, and (b) a gas of particles described by a Boltzmann equation.  However, the particular configurations they investigated do not show the effects that are found here.  That is because
their analyses were done in local coordinates, in terms of weak-field plane waves, and so the non-local effects, and the transit of the matter across the sphere of directions outwards from the source, do not appear.
Also, none of these works considered the radiation-dominated regime.

Finally, there has been much work on the scattering of light and radio signals, in the geometric-optics limit, by gravitational waves.  However, on one hand, this has very largely dealt with coordinate, rather than invariant, computations of scattering; and, on the other, the intended applications of this to look for gravitational-wave modulations of signals from astrophysical sources are only sensitive to differential effects (one must compare two signals, neighboring in time or space) \cite{ADH2013}.

\subsection{Organization}

Section \ref{Prel} reviews the Bondi--Sachs asymptotics which will be used.  Section III establishes the key relation between gravitational radiation and asymptotic holonomy.  In Section IV the equations governing the asymptotically constant fields are derived; these are used to get the basic formulas for energy--momentum exchanges in Section V.
Section VI goes over the relation to linearized theory, which involves a fine point.  Section VII discusses the response of test particles, giving in particular general formulas for scattering of them in the case of linearized quadrupolar waves.
I pointed out above that in many cases the waves' high frequencies (compared to the matter's dynamical time-scales) leads to an averaging-out of energy--momentum transfers; however, Section VIII discusses three classes of cases in which non-zero average effects are possible.  The final section contains a brief summary and discussion.

{\em Notation and conventions.}  The notation and conventions are those of Penrose and Rindler \cite{PR1986}, except where explicitly indicated.  These books also serve as a reference for all material not otherwise explained, including the spin-coefficient calculus in the form given by Geroch, Held and Penrose \cite{NP1961,GHP}.
The metric signature is $+{}-{}-{}-$, and the curvature tensors satisfy $[\nabla _a,\nabla _b]v^d=R_{abc}{}^dv^c$ and $R_{ac}=R_{abc}{}^b$.  The speed of light $c$ is often suppressed.  Einstein's equation (without cosmological constant) is $R_{ac}-(1/2)Rg_{ac}=-8\pi (G/c^2) T_{ac}$.
A familiarity with two-component spinors is assumed for some of the derivations (the treatment without them is significantly more labored), but the main results are given in tensor form.

\section{Preliminaries}\label{Prel}

We recall here the main elements of the Bondi--Sachs asymptotics, and their expression in terms of Newman--Penrose spin-coefficients.  The main point which will be used explicitly is the Sachs peeling property.  However, because the nonlocality of the gauge plays such an important implicit role, we also review how this arises.  The reader familiar with these points can skip this section.  

All of this material can be found in ref. \cite{PR1986}, and no proofs are given. 

{\em Asymptotic hypotheses.}
We may say that a space--time {\em admits Bondi--Sachs--Penrose} asymptotics if:
(a) the space--time $(M,g_{ab})$ embeds as the interior of a manifold with boundary $\hat{M}=M\cup\scrif$, where $\scrif =\partial\hat{M}$; (b) there is a non-negative function $\Omega :\hat{M}\to\R$ of class $C^3$ with $\Omega$ vanishing precisely on $\scrif$ but ${\hat\nabla}_a\Omega$ nowhere zero on $\scrif$; (c) the rescaled metric ${\hat{g}}_{ab}=\Omega ^2g_{ab}$ is Lorentzian and $C^3$ on $\hat{M}$; (d) all matter fields vanish at $\scrif$ (that is, the stress--energy $T_{ab}$ has a well-defined limit of zero at $\scrif$) and the cosmological constant $\lambda =0$; (e) each point on the boundary $\scrif$ is a future (but not a past) end-point of null geodesics in $(M,g_{ab})$; (f) the boundary $\scrif$ is a ${\hat{g}}_{ab}$-null hypersurface diffeomorphic to $S^2\times\R$, and the $\R$ factors can be taken to be ${\hat{g}}_{ab}$-null generators.

A few comments are in order.  First, the assumption that the cosmological constant is zero means that cosmological effects are not important over the scale of the isolated system we are modeling; it is not a cosmological hypothesis.  Second, the assumptions are not all independent; they have been included for convenience.  Third, these assumptions are very nearly those of weak future asymptotic simplicity, but for reasons explained elsewhere \cite{ADH2011} I prefer to not to rely on some of the hypotheses of that concept.

{\em Bondi coordinates and nonlocality.}
With these assumptions, we may introduce a Bondi coordinate system $(u,r,\theta,\varphi )$ in a neighborhood of $\scrif$.  Here $u$ is a null coordinate, the Bondi {\em retarded time}, with the $u=\const$ hypersurfaces meeting $\scrif$ transversely.   Since $u$ is null, these hypersurfaces are ruled by null geodesics with parallel-transported null tangent $l^a=\nabla ^au$; this defines the {\em outgoing null congruence} associated with the Bondi system.  
We take $r$ to be an affine parameter along the geodesics of this congruence, normalized by $l^a\nabla _a r=1$.
The zero of $r$ may be set by a natural device; see e.g. \cite{PR1986}.  We may take $\Omega =1/r$.  The regularity of the rescaled metric, and the hypotheses on the curvature, at $\scrif$ then imply certain asympotic expansions in $r$.

The angular coordinates $(\theta ,\varphi )$ label the generators of $\scrif$, and these extend to coordinates on space--time by
mapping a point in space--time to the $(\theta ,\varphi )$ values of the end-point of the member of the outgoing congruence through the point.
However, at this point we know only that the $r=\const$, $u=\const$ surfaces are diffeomorphic to spheres, so the angular coordinates are as yet determined only up to a diffeomorphism of $S^2$.  It is in restricting this freedom that the nonlocal gauge choice, which is ultimately responsible for the definition of the asymptotically constant vectors, enters.

One can show that the metrics on the $u=\const$, $r=\const$ surfaces have a well-defined $u$-independent conformal structure as $r\to\infty$.  (One shows that with the asymptotic hypotheses, the shear up the generators of $\scrif$ vanishes.)  These surfaces must then, up to a conformal factor, approach ordinary metric spheres (or, more properly, spheres with the negatives of the ordinary metric).  The conformal factor is $\Omega ^2=r^{-2}$ gives a well-defined metric on the surfaces (by (c)), so we must be able to choose coordinates $(\theta ,\varphi )$ such that $\Omega ^2g_{ab}$, when restricted to vectors tangent to these surfaces, approaches $-(d\theta ^2+\sin ^2\theta d\varphi ^2)$.  
The construction of these coordinates --- equivalent to finding a complex stereographic coordinate on the sphere --- implicitly involves the solution of an elliptic partial differential equation on the sphere.  This is a nonlocal problem.  Indeed, the nonlocality is much stronger than, for example, that involved in Newtonian potentials, for here the sphere represents the family of all asymptotic null directions.  In other words, this nonlocality does not fall off with distance from the source; it reflects the problem of correlating (even asymptotically distant) frames at different angles around the source.  In this sense it is scale-free.

If the conformal structure and orientation of a sphere are known, then the structure present is that of a Riemann sphere, and the symmetries are the fractional linear transformations $SL(2,\C)/\{ \pm I\}$.  This group is of course isomorphic to the proper orthochronous Lorentz group, and the sphere has naturally the structure of the light rays through an event in Minkowski space.  The choices of unit sphere metrics on it compatible with this structure are in one-to-one correspondence with the choices of a unit timelike vector in Minkowski space, which would allow the identification of the set of light-rays with a unit sphere at time coordinate unity.  Applying this now to the $u=\const$ spheres at $\scrif$, we see that the choice of unit sphere metric can be thought of as a choice of asymptotic time direction.  In fact, it turns out that there is a well-defined way of comparing these directions at different $u$-values, and in a Bondi system we restrict the allowable choices of $u$ so that these time directions are all the same.
Once the choice of time-direction has been made, there remains an $SO(3)$ freedom in the choice of $\theta$ and $\varphi$.

{\em The null tetrad.}
We now introduce a null tetrad $l^a$, $m^a$, ${\overline m}^a$, $n^a$ compatible with the Bondi system.  We keep $l^a$ the parallel-transported null vector along the outgoing congruence, and we require $m^a$ and $n^a$ to be transported parallel along $l^a$.  We also require $n^a$ to have a well-defined limit at $\scrif$ where it becomes tangent to the null generators of $\scrif$, and normalized so that $n^a\nabla _au=1$.  (In fact, for the standard null tetrad, we require slightly more, namely that ${\tilde l}^b{\tilde\nabla}_bn^a$ vanish at $\scrif$, where ${\tilde l}^a=\Omega ^{-2}l^a$ off $\scrif$ and ${\tilde l}^a$ is defined by continuity on $\scrif$.)
Then $m^a$ will lie tangent to the $u=\const$ hypersurfaces.  We will also use an associated spinor dyad $\omicron ^A$, $\iota ^A$, so $l^{AA'}=\omicron ^A\omicron ^{A'}$, $m^{AA'}=\omicron ^A\iota ^{A'}$, $n^a=\iota ^A\iota ^{A'}$.  We shall not need a definite choice of phase for $m^a$ --- such a choice would conventionally be associated with a particular choice of $(\theta ,\varphi )$.

{\em Asymptotic expansions.}
The spin-coefficient equations can be integrated inwards along the congruence defined by $l^a$ to get asymptotic expansions for the tetrad, the spin-coefficients, and the curvature components.

The null tetrad vectors have the coordinate forms
\begin{eqnarray}
l^a&=&\partial _r\label{tetl}\\
m^a&=&(r\sqrt{2})^{-1}\left(\partial _\theta +i\csc\theta \partial _\phi \right)\nonumber\\ &&+
\left(O(r^{-2})\mbox{ terms in }\partial _\theta\, ,\ \partial _\varphi\right)
  +\omega \partial _r\label{tetm}\\
n^a&=&\partial _u+U\partial _r+X^\theta\partial _\theta +X^\varphi \partial _\varphi\label{tetn}
\end{eqnarray}
where $\omega$, $U$, $X^\theta$, $X^\varphi$ are functions with the asymptotic properties
\begin{equation}
\omega =O(r^{-1})\, ,\ U=-1/2+O(r^{-1})\, ,\ X^\theta, X^\varphi =O(r^{-3})
\end{equation}
along the outgoing null congruence.  Here the symbol $O(g)$ means a term whose magnitude is bounded by an $r$-independent multiple of $|g|$ as $r\to\infty$.  These expansions hold uniformly in the angular directions and locally uniformly in $u$.
(Again, near any given space--time point one can use the freedom in choosing $(\theta ,\phi )$ to avoid coordinate singularities.)

We note for later use that the tangents to a $u=\const$, $r=\const$ surface are $m^a-\omega l^a$, ${\overline m}^a-\overline\omega l^a$.  These are evidently normalized (and differ from $m^a$, ${\overline m}^a$ only by a null rotation).  Thus the surface area element is $(2i)^{-1}(m^a-\omega l^a)\wedge ({\overline m}^b-\overline\omega l^b$).

{\em Sachs peeling.}
The Weyl tensor is $C_{abcd}=\Psi _{ABCD}\epsilon _{A'B'}\epsilon _{C'D'}+{\overline\Psi}_{A'B'C'D'}\epsilon _{AB}\epsilon _{CD}$, and its components with respect to the dyad are $\Psi _n$ (that is, the spinor $\Psi _{ABCD}$ contracted with $n$ iotas and $4-n$ omicrons).  

A key consequence of the assumptions (a)--(f) is the {\em Sachs peeling property}
\begin{equation}
\Psi _n=\frac{\Psi _n^0(u,\theta,\varphi)}{r^{5-n}}+O(1/r^{6-n})\, .
\end{equation}
Note in particular that $\Psi _4$, which represents the part of the field transverse to the outgoing null congruence, falls off as $1/r$ and the other components fall off more rapidly.  Thus $\Psi _4$ is referred to as the {\em radiative} part of the field. 
It is linked by the asymptotic Bianchi identity
$\teth\Psi _4^0=\partial _u\Psi _3^0$ to $\Psi _3^0$, and this latter component is sometimes called the {\em semiradiative} part.  
The {\em Bondi news} $N=N(u,\theta ,\phi )$ is a potential for these:  one has $\Psi _3^0=\teth N$, $\Psi _4^0=\partial _uN$.  
Here $N=-\partial _u{{\overline\sigma}}^0$, with
\begin{equation}
\sigma =\sigma ^0/r^2+O(r^{-3})
\end{equation} the asymptotic expansion of the shear.  Two other
spin-coefficients'
asymptotic forms involving the radiation will be important:
\begin{equation}
  \sigma '=N/r+O(r^{-2})\, ,\quad
    \kappa '=\teth N/r+O(r^{-2})\, .
\end{equation}

{\em Bondi--Metzner--Sachs group.}
The group of transformations preserving the asymptotic structure is the {\em Bondi--Metzner--Sachs} group.  It can be viewed as the group of coordinate transformations preserving the Bondi--Sachs form of the metric.  It is the semidirect product of the (proper, orthochronous) Lorentz group acting on the asymptotic sphere of directions with the {\em supertranslations} $u\to u+\alpha (\theta ,\varphi )$.  

\section{Gravitational radiation and asymptotic geometry}\label{grag}

In order to get to the main ideas as rapidly as possible, we will begin, in subsection~A, by showing how Sachs peeling gives rise to the holonomic obstructions 
which are at the heart of the scattering.  While the formulas are simple, we will find that to interpret them rigorously we need to define a class of tensor and spinor fields in the asymptotic regime scaling in a different way from those often used in the $\scrif$ formalism; we call these the {\em physically bounded} fields and treat them in subsection~B.
Subsection C, which may be skipped, discusses the conditions for radiation-dominance and gives some estimates for astrophysical situations in which they might hold.

\subsection{Gravitational radiation and holonomy}

It is commonly asserted that isolated radiating  general-relativistic
systems are modeled by space--times which are asymptotically flat.
This, however, is not entirely true, and accepting it too uncritically
would lead to missing key physical features of these systems.  It is
in fact a {\em signature}  of gravitational radiation that the effects
of curvature are stamped on the asymptotic geometry as finite effects,
even in the limit of passage to infinitely distant regions.

This follows directly from the scaling of the radiative and semiradiative parts of the field according to Sachs peeling.
Consider the increment a vector receives on
being transported parallel around an area element $\rmd S^{pq}$, which
is given by the holonomy $R_{pqb}{}^a\rmd S^{pq}$. If we take $\rmd
S^{pq}$ to be determined by an interval $\delta u$ in Bondi retarded
time and a change $\delta \mu=(\delta\theta -\rmi\sin\theta\delta\phi) /
\sqrt{2}$ in angle, then 
to leading order in $r$, according to (\ref{tetl})--(\ref{tetn}), we will have
$\delta u\,\partial _u=\delta u (n^a+(1/2)l^a)$ but
\begin{equation}
 \delta\theta\,\partial _\theta +\delta\phi \,\partial _\phi 
 =r(\delta\mu \, m^a+\overline{\delta\mu} \, {\overline m}^a)
+\text{a multiple of }l^a\, ,
\end{equation} 
where the $r$ factor arises because the change $\delta\mu$ in angle gives a physical displacement scaling as $r$.
On the other hand, the gravitational radiation field, the
leading component of $R_{pqa}{}^b$ as $r\to\infty$, falls off as
$1/r$.  One is thus left with a {\em finite} holonomy even as
$r\to\infty$, which one can check is 
\begin{equation}\label{asprop}
  \delta u\delta\mu \Psi _4^0(u,\theta ,\phi )\, (l_am^b-m_al^b)
+\mbox{conjugate}\, .
\end{equation}
Sachs peeling of the radiative term 
balances the physical scaling due to a change in asymptotic angle, leaving one with a finite effect.  

One finds, similarly, that for $\rmd S^{ab}$ spanning the two
independent angular directions that there is finite limiting holonomy
proportional to $\Psi _3^0(u,\theta ,\phi ) (l_am^b-m_al^b)$ (and
conjugate). Recall that $ \Psi _4^0$, $\Psi _3^0$ exactly measure the
gravitational radiation content of the field (and are linked to each
other by an asymptotic Bianchi identity).

The analysis
here has dealt with parallel transport in three of the four
dimensions:  changes in angle and in retarded time.  One can also
consider the remaining direction, that is, passage outward to more
distant regions.  
The leading contribution is due to holomomies spanned by $\delta r\,\partial _r=l^a$ and $\delta\theta\,\partial _\theta +\delta\phi\,\partial _\phi$ is proportional to $\delta r\delta\mu\, \Psi _2^0/r^2$ (and conjugate) and to those spanned by $\delta r\, \partial _r$ and $\delta u\,\partial _u$ is $\delta r\delta u\Psi _3^0/r^2$ (and conjugate).  Thus, even integrating outwards from $r$ to infinity, the holonomies they contribute will have dominant terms bounded by $\delta\mu\, \Psi _2^0/r$ and $\delta u\Psi _3^0/r$ (and conjugates).
Both of these vanish as $r\to\infty$, so they contribute no holonomy in the limit. 
In other words, the contributions to the ambiguities in identifying vectors at different points due to propagation outwards along $l^a$ vanish as $r^{-1}$.

In these arguments, two sorts of scalings in the asymptotic regime are used.  First, we wish to compare what happens as we go out to very great distances along different outgoing null directions; it is in studying this that the relevant displacement field $r(\delta\mu\, m^a+\delta\overline\mu\, {\overline m}^a)$ has the factor of $r$, indicating the increasing distance in physical space--time corresponding to a fixed change in asymptotic angle.  On the other hand, in maintaining that the limit (\ref{asprop}) is finite we imply that we do not need to worry about factors of $r$ in applying (\ref{asprop}) to vectors or covectors.  In fact, to make precise what the interpretation of the limit (\ref{asprop}) is, we must specify what objects it acts on.

We shall do this formally in the next subsection.  However, if an asymptotically {\em covariantly} constant vector field did exist, we should certainly expect its components with respect to the tetrad to be bounded, and evidently (\ref{asprop}) then provides an obstruction. 
We see then that {\em precisely} for gravitationally radiating
space--times, there are no asymptotically covariantly constant vector
fields: curvature obstructions to their existence persist as finite
limits as one passes to future null infinity.

\subsection{Physically bounded fields}

The underlying reason the $\scrif$ formalism is so useful is that, in many cases of interest, the appropriate scaling of physical quantities with the affine parameter $r$ turns out to be equivalent to the extensibility of the quantity to the conformal boundary.  Conversely, geometrically natural structures on the conformally rescaled manifold with boundary ${\hat M}=M\cup\scrif$ have certain space--time asymptotic scalings.  However, some of those which have been most commonly used are not well-adapted to the questions here.  

A vector field defined in the neighborhood of some point on $\scrif$ will in general, in the physical space--time, appear to diverge in the physical space--time as one approaches $\scrif$ along the outgoing Bondi--Sachs congruence.  (This divergence is with respect to parallel-propagation, and also in the Bondi coordinates.)  This is because, starting from a finite point, one need only flow by a finite parameter to reach $\scrif$ along the field.

In this paper, we shall be concerned with spinor and tensor fields which are candidates for being, in a sense to be made precise, asymptotically constant.  Such fields should have bounded components with respect to the Bondi dyad or tetrad as $r\to +\infty$.\footnote{Note that in such statements the points in question have angular coordinates in the interior of the chart, that is, the chart is chosen so that we avoid the polar coordinate singularities.}
We would expect, based on the
argument of the previous paragraph, that these fields must have no components transverse to $\scrif$.  This is indeed the case, as we now make precise.

We shall work in terms of spinors, from which the results for other quantities can be obtained.  If $\omicron ^A$, $\iota ^A$ is a normalized dyad adapted to the Bondi system, then the rescaled dyad $\tomicron ^A=\Omega ^{-1}\omicron ^A$, $\tiota ^A=\iota ^A$ has a non-zero limit on $\scrif$.  
(And dually $\tomicron _A=\omicron _A$, $\tiota _A=\Omega\iota _A$.)
Thus any spinor field $\xi ^A$ can be expressed as
\begin{eqnarray}
  \xi ^A&=&\xi ^0\omicron ^A+\xi ^1\iota ^A\\
    &=&\txi ^0\tomicron ^A +\txi ^1\tiota ^A
\end{eqnarray}
with
\begin{equation}
  \txi ^0=\Omega \xi ^0\, ,\ \txi ^1=\xi ^1\, .
\end{equation}
We will say the field is {\em physically bounded} if $\xi ^0$, $\xi ^1$ are bounded as $r\to\infty$.  Note that it makes sense to speak of the degree of differentiability of a physically bounded field, as a field on $\scrif$.  We will here be interested in physically bounded fields which are at least $C^2$ (because we want to be able to differentiate them once in a direction transverse to $\scrif$ and once in a direction tangential to it).      

While the characterization of physically bounded fields has been given with respect to a dyad, it is easy to see it is Bondi--Metzner--Sachs invariant.  For it is equivalent to requiring $\tiota _A\xi ^A$ to vanish at $\scrif$, and $\tiota _A$ is invariantly defined on $\scrif$, up to proportionality.

From a physical point of view, the limiting components of the field are those with respect to the physical, rather than the rescaled dyad.  We have $\xi ^1=\txi ^1$, but we have $\xi ^0=-{\tilde l}^a{\tilde\nabla}_a\txi ^0=-{\tilde\thorn}\txi ^0$ in the limit, since ${\tilde\thorn}\Omega =-1$ at $\scrif$.  Note that these limiting values are invariant under
supertranslation-induced changes of dyad (since under a supertranslation $\tiota ^A$ is preserved and $\tomicron ^A$ changes by a multiple of $\tiota ^A$).

At a given point of $\scrif$, the set of all limiting values of physically bounded spinor fields evidently forms a two-complex-dimensional vector space, and the family of these as the point on $\scrif$ varies forms the bundle of physically bounded spinors over $\scrif$, which can be viewed as a boundary of the spin-bundle over the physical space--time.  This bundle is {\em not} the spin bundle over the conformally rescaled $(\hat{M}, {\hat g}_{ab})$.  However, it is a direct sum of certain spin- and boost-weighted bundles there, for $\xi ^1=\xi ^A\tomicron _A$ is a section of the bundle of Geroch--Held--Penrose type $\{ 1,0\}$, and $\xi ^0$ a section of the bundle of type $\{ 0,1\}$, over $\scrif$ with respect to the rescaled metric.

Higher-valence physically bounded fields are defined as tensor products of the physically bounded spinor fields and their duals.  

These definitions allow us to interpret the arguments of the previous subsection rigorously; we see that (\ref{asprop}) precisely defines a holonomy on the space of physically bounded vectors at $\scrif$.

\subsection{Some estimates for radiation-dominance}

The main idea in this paper is that the Bondi--Sachs treatment of energy--momentum, extended to the radiation zone, resolves the holonomic obstructions.  This in turn depends on the holonomy being well-approximated by the lead terms in the Bondi--Sachs expansion, in other words, the effects of any matter there being relatively negligible, what I called {\em radiation-dominance}.
In this subsection, I give some rough estimates for this condition to be fulfilled.  The aim is not to be exhaustive, but to give the reader a sense of the scales involved.  The results here are not used elsewhere in this paper; this section may be skipped.

The essential restrictions are:  that $\Psi _3$ and $\Psi _4$ should be well-modeled by their leading asymptotic forms; and that the other curvature terms contribute negligible amounts, in the sense of their actions by holonomy on physically bounded quantities.  In fact, the physical quantities of interest in any given situation will not depend on every detail of these curvature components but only on certain integrals, and thus one really should have an averaging scale in mind.  However, here we will (conservatively) look at the full infinitesimal holonomies.

The question of just when (for which values of $r$) the components $\Psi _3$ and $\Psi _4$ take on their asymptotic forms depends on the details of the system.  Roughly speaking we expect that for a wave-packet of nominal angular frequency $\omega$ this will occur once $r\gg c/\omega$.  (This corresponds the frequency-based definition of a wave zone used by some authors.)  However, there will also be emissions of radiation which are not well-represented by wave-packets, because they involve ``memory effects'' with arbitrarily low-frequency components.  For instance, if such an effect is due to ejection of a jet, one expects the radiation zone to take over sufficiently far from the jet (a moving boundary, with $r$ increasing with time).

I will now turn to the restrictions on the matter (although, for conceptual clarity, I will keep track of the Weyl tensor components as well).
These conditions are estimated
by considering holonomies as in Section III.A. Suppose we require a precision $\eta _{\rm ang}$ for the generators of holonomies in the angular directions (spanned by $rm^a$, $r{\overline m}^a$).  Then we
find
\begin{equation}\label{angrest}
  |r^2\Psi _1|\, ,\   |r^2\Psi _2|\, ,\ 
  |r^2\Psi _3-\Psi _3^0|\, ,\  
  | r^2\Phi _{1q}|\, ,\ 
  |r^2\Lambda |<\eta _{\rm ang}
\end{equation} 
(where $R_{ab}=6\Lambda g_{ab}-2\Phi_{ab}$ with $\Phi _{ab}$ trace-free) for $q=0,1,2$.  
For a requisite precision $\eta _{u,\,{\rm ang}}$ for infinitesimal
holonomies in an angular direction and the $u$-direction
\begin{equation}\label{urest}
  |r\Psi _2|\, ,\ 
  |r\Psi _3|\, ,\ 
  |r\Psi _4 -\Psi _4^0|\, ,\ 
  |r\Phi _{2q}|<
  \eta _{u,\,{\rm ang}}\, ,
\end{equation}
with $q=0,1,2$.
(Note that these quantities have units inverse time, in contrast to those from the angular-angular directions.)
There are also stability restrictions, for holonomies involving the outward direction, but these are very similar and will not be discussed explicitly.

What sorts of sensitivities $\eta _{\rm ang}$, $\eta _{u,\, {\rm ang}}$ are needed?  The holonomy we wish to detect in angular directions is $\teth N$, and in $u$-angular directions $\partial _u N$.  Thus the sensitivities required can be suggestively written as
$l\delta N$, $\omega \delta N$ where 
$\delta N$ is a measure of the sensitivity sought for the news, and $l$, $\omega$ are measures of the effective spherical harmonic, and effective angular frequency, of the news.  (These are simply notations to help understand what is going on --- the source need not be a pure spherical harmonic, nor purely monochromatic.)  
We have then
\begin{equation}
\eta _{\rm ang}=\delta N/l\, ,\quad
\eta _{u,\ {\rm ang}} =\omega\delta N\, .
\end{equation}
Note that in the radiation zone we expect to have $r\gg c/\omega$, and hence
\begin{equation}
  (r/c)\eta _{u,\ {\rm ang}}\gg l\eta _{\rm ang}
>\eta _{\rm ang}\, .
\end{equation}
This means that where the restrictions (\ref{angrest}), (\ref{urest}) overlap, it is the former which is the more stringent.\footnote{If memory effects are to be considered, some elements of the discussion in this paragraph must be modified.}

To get an idea of the scales involved, let us consider as a source a binary of equal masses, each $M$, in non-relativistic mutual circular orbits with orbital angular frequency $\omega$ (so the angular frequency of the gravitational waves is $2\omega$).  Then we have
\begin{eqnarray}
  |N|&\simeq& 8 (GM\omega /c^3)^{5/3}\nonumber\\
&=& .04\times \left( \frac{M}{1.4M_\odot}\cdot\frac{2\omega /(2\pi)}{ 10^3\, {\rm s}^{-1}}\right) ^{5/3}\, ,
\end{eqnarray}
where $M_\odot$ is the Sun's mass and the figures are scaled to correspond to a late-stage binary of neutron stars.  For $2\omega/(2\pi )=1\, {\rm s}^{-1}$, the estimate would be $4\times 10^{-7}$.  
Thus if we would like to measure radiation from a double neutron binary with wave frequency about $1\, {\rm s}^{-1}$, it would be reasonable to take $\eta _{\rm ang}=10^{-8}$ or so; we could get by with $\eta _{\rm ang}=10^{-3}$ in the later stages.

We can determine from this restrictions on the radiation zone.
Taking $\Psi _2\simeq 2.8(GM_\odot /c^2)/r^3$, we see from (\ref{angrest}) that we must have $r\gtrsim 2.8 (GM_\odot /c^2)/\eta _{\rm ang}$.  This would be $2\times 10^{13}\, {\rm cm}$ (roughly $1\, {\rm au}$) for $\eta _{\rm ang}=10^{-8}$.  
(The constraints from $\Psi _1$ and $\Psi _0$ would be expected to be weaker.)  This marks the onset of the radiation zone, in the absence of matter.

What if matter is present?  If the matter is non-relativistic, the most severe constraints will come from the $\Phi _{11}$ terms; we will have $\Phi _{11}\sim 2\pi G \rho /c^2$, with $\rho$ the density. Thus we will have $\rho r^2\lesssim \eta _{\rm ang}c^2 /(2\pi G )$.  To put in some standard astrophysical scales, this may be cast as
\begin{equation}
\left( \frac{\rho}{10^{-24}\, {\rm g}\, {\rm cm}^{-3}}\right)
\left(\frac{r}{10^{13}\, {\rm cm}}\right)^2 \lesssim 2\times 10^{17}\left(\frac{\eta _{\rm ang}}{10^{-8}}\right)\, .
\end{equation}
So, for instance, for $\eta _{\rm ang}=10^{-8}$, radiation-dominance should apply for a non-relativistic medium of density $10^{-24}\, {\rm g}\, {\rm cm}^{-3}$ up to a radius of about $4\times 10^{21}\, {\rm cm}\simeq 1\, {\rm kpc}$.

These examples show that the condition of radiation-dominance, as defined in this paper, can apply in realistic situations.

\section{Asymptotically constant fields}\label{acv}

We have seen that, when gravitational radiation is present, there will be no covector fields which are asymptotically covariantly constant.  
Nevertheless, there is a space of asymptotically constant covectors $T^*$ used in the construction of the Bondi--Sachs energy--momentum.  While the starting-point for this is a set of covector fields, the results are not covector fields in the ordinary sense, because components are discarded and limits are taken.

The goal here is to define asymptotically constant fields in a neighborhood of $\scrif$.  It is simplest to make the construction for spinors; other fields are then determined from the tensor algebra of those.  The first subsection gives a treatment of the fields at $\scrif$; by using physically bounded fields we recover all the relevant components there.  The second subsection extends the construction inwards, to the physical space--time.

\subsection{Asymptotic constancy at null infinity}

An asymptotically constant spinor field will be physically bounded.   If $\xi ^A$ is a physically bounded field which is a candidate for asymptotic constancy, then we shall want to examine its physical covariant derivative $\nabla _a\xi ^B$ in various directions in the limit that the points in question approach $\scrif$.  If we investigate the directions tangent to $\scrif$ (the behavior transverse to $\scrif$ will be the focus of the next subsection), then we are interested in $\nabla _a\xi ^B$ contracted with the physical dyad $\omicron _B$, $\iota _B$ but the rescaled tetrad vectors ${\tilde n}^a$, ${\tilde m}^a$, ${\tilde{\overline m}}^a$, as the points approach $\scrif$.  This amounts to asking for the
behavior of $\nabla _a\xi ^B$ as a one-form on $\scrif$ with values in the physically bounded spinors. 
Because of the holonomic obstructions, requiring this quantity to vanish is too strong.  One keeps only certain components of it.

Recall that the
physical dyad $\omicron ^A$, $\iota ^A$ rescales
according to $\tomicron ^A=\Omega ^{-1}\omicron ^A$, $\tiota ^A=\iota
^A$ to achieve finite limits at $\scrif$ as spinor fields on the conformally rescaled space--time,
and so
\begin{equation}
 \txi ^0=\Omega\xi ^0\mbox{ and }\txi ^1=\xi ^1\, .
\end{equation}
We shall not distinguish between $\txi ^1$ and $\xi ^1$, nor between
$\tio ^A$ and $\iota ^A$.
This means that $\txi
^0$ will vanish at $\scrif$; in the usual treatment of Bondi--Sachs
asymptotic constancy, done strictly at $\scrif$, the field $\txi ^0$
is simply omitted, and the definitions cast entirely in terms of $
\txi ^1=\xi ^1$.   

The usual definition of Bondi--Sachs constancy is equivalent to
requiring that $\tiota ^{A'}\nabla _{AA'}\xi ^B$ vanish on $\scrif$ when contracted with the {\em rescaled} dyad.  Using the relation
\begin{equation}\label{conresc}
 \nabla _{AA'}\xi ^B=\tnab _{AA'}\xi ^B-\epsilon _A{}^B\Upsilon
_{CA'}\xi ^C
\, ,
\end{equation}
where $\Upsilon _a=\Omega ^{-1}\tnab _a\Omega =-\Omega ^{-1}\tn _a$, 
we find,
after a brief calculation, that the Bondi--Sachs constancy condition 
is
\begin{equation}
 \teth\xi ^1\approx 0\quad\mbox{and}\quad{\tthorn}'\xi ^1\approx 0
  \, ,
\end{equation}
where $A\approx B$ means $A$ and $B$ are equal at $\scrif$.
(In this computation, and the ones which follow, the only issue which requires a bit of work is the limit associated with the vanishing of $\Omega$ in $\Upsilon _a$.  Note that each side of eq. (\ref{conresc}) is independent of the choice of conformal factor with the allowed class, and so will be the vanishing of the components in question.  The most direct way of doing the limit, keeping with this paper's general formalism, is to use the standard asymptotic expansions for the spin-coefficients in the Bondi--Sachs frame (\cite{PR1986}, pp. 394--395), taking $\Omega =1/r$.)
Again, this 
system makes no mention of the field $\xi ^0$, and
that field is not generally used in analyses on $\scrif$.

In order to fix the field $\xi ^0$, we shall require that 
$n^a\nabla _a\xi^B$, ${\tilde m}^a\nabla _a\xi ^B$
should vanish at $\scrif$ as 
physically bounded spinors.  
We have
\begin{eqnarray}\label{neq}
  n^a\nabla _a\xi ^B&=&\Omega \tom ^B\tthorn '\xi ^0 +\iota ^B\tthorn
'\xi ^1 +\Omega\xi ^0\tthorn '\tom ^B+\xi ^1\tthorn '\iota ^B\qquad\nonumber\\
&=&\Omega \tom ^B\tthorn '\xi ^0 +\iota ^B\tthorn
'\xi ^1 -\Omega\xi ^0\tilde{\tau} \iota ^B-\xi ^1{\tilde{\kappa}}'\tom 
^B\nonumber\\
&=& \omicron ^B\tthorn '\xi ^0 +\iota ^B\tthorn
'\xi ^1 -\Omega\xi ^0\tilde{\tau} \iota ^B-\xi ^1\Omega ^{-1}{\tilde{\kappa}}'\omicron
^B
 \, .\qquad
\end{eqnarray}
We recall that the values of the components of this are defined to be
the coefficients of $\omicron ^B$ and $\iota ^B$ at $\scrif$.  Now
${\tilde\kappa}'$ vanishes at $\scrif$, and therefore
$\Omega ^{-1}{\tilde\kappa}'=-\tthorn{\tilde\kappa}'$ to first order at $\scrif$.
However, one of the spin-coefficient equations gives us
$\tthorn{\tilde\kappa}'\approx 0$.  Therefore the vanishing of
the components of eq.~(\ref{neq}) is equivalent to
\begin{equation}
 \tthorn '\xi ^0 \approx 0\, ,\qquad \tthorn '\xi ^1\approx 0\, .
\end{equation}
Similarly, we 
have
\begin{eqnarray}
\tm ^{AA'}\nabla _{AA'}\xi ^B&=&
  \Omega \tom ^B\teth\xi ^0+\iota ^B\teth\xi ^1 +\Omega\xi ^0\eth\tom
^B +\xi ^1\eth\iota ^B\qquad\nonumber\\
&=&
  \Omega \tom ^B\teth\xi ^0+\iota ^B\teth\xi ^1 -\Omega\xi
^0\tilde{\sigma}\iota ^B -\xi ^1{{\tilde\rho}'}\tom ^B\nonumber\\
&=&
  \omicron ^B\teth\xi ^0+\iota ^B\teth\xi ^1 -\Omega\xi
^0\tilde{\sigma}\iota ^B -\xi ^1\Omega ^{-1}{{\tilde\rho}'}\omicron ^B
\, .\qquad
\end{eqnarray}
We have ${\tilde\rho}'\sim 1/(2r)$, and so ${\tilde\rho}'\approx 0$, $\Omega ^{-1}{\tilde\rho}'\approx 1/2$.
Thus we obtain
\begin{equation}
\teth\xi ^0-(1/2) \xi ^1\approx 0\, ,\qquad
  \teth\xi ^1\approx 0\, .
\end{equation}

We may therefore collect our equations for an asymptotically constant
spinor field, in terms of data at $\scrif$, as
\begin{equation}\label{beq}
\tthorn '\xi ^0\approx 0\, ,\ \tthorn '\xi ^1\approx 0\, ,\ 
  \teth\xi ^0 -(1/2)\xi ^1\approx 0\, ,\  \teth\xi ^1\approx 0\, .
\end{equation}
It follows from
standard spin-coefficient formulas that these equations are integrable
and have a two-complex-dimensional space of solutions; also, these
equations imply
\begin{equation}\label{beqa}
\teth '\xi ^0\approx 0\, ,\ \teth '\xi ^1\approx -\xi ^0\, .
\end{equation}
{\em We take equations~(\ref{beq}),~(\ref{beqa}) to determine the 
asymptotically 
constant spinors.}  Asymptotically constant
vectors, covectors, etc., are determined from these.

We compute for use below the remaining derivative, tangent to $\scrif$, of a physically bounded field:
\begin{eqnarray}\label{mbeq}
{\tilde{\overline m}}^a\nabla _a\xi ^B
&=&{\tilde{\overline m}}^a{\tilde\nabla}_a\xi ^B
-\iota ^B\Omega ^{-1}{\tomicron}^{A'}\nabla _{CA'}\xi ^C\nonumber\\
&=&{\tilde\eth}'\xi ^B -\iota ^B\Omega ^{-1}
\left( {\tilde\xi}^0{\tilde\thorn}\Omega +\xi ^1{\tilde\eth}'\Omega\right)\nonumber\\
&=&\left( {\tilde\eth}'\xi ^0 -\Omega ^{-1}{\tilde\sigma}' \xi ^1 \right)\omicron ^B\nonumber\\
&&+\left({\tilde\eth}'\xi ^1 -{\tilde{\overline\rho}}'{\tilde\xi}^0
  -\Omega ^{-1}\left( {\tilde\xi}^0{\tilde\thorn}\Omega +\xi ^1{\tilde\eth}'\Omega\right)\right)\iota ^B\nonumber\\
&\approx&\left( {\tilde\eth}'\xi ^0+{\dot{\overline\sigma}}^0\xi ^1\right)\omicron ^B
+\left({\tilde\eth}'\xi ^1 +\xi ^0\right)\iota ^B\, .
\end{eqnarray}
(The reader used to the calculus in terms of the rescaled quantities should note that the symbol $\approx$ at the last step is here used for equality at $\scrif$ of physically bounded fields.)

Now suppose that $\xi ^A$ is asymptotically constant.  Then, as a one-form on $\scrif$ with values in the physically bounded fields, we have, using eqs. (\ref{beqa}), (\ref{mbeq}),
\begin{eqnarray}
\nabla _a\xi ^B
  &=&-\tm _a{\tilde{\overline m}}^c\nabla _c\xi ^B\nonumber\\
&=&-\tm _a{\dot{\overline\sigma}}^0\omicron _D\omicron ^B\xi ^D\, .\label{nabxi} 
\end{eqnarray}
(As a one-form on $\scrif$, only the contractions of this with $\tn ^a$, $\tm ^a$, ${\tilde{\overline m}}^a$ are determined; we could add to this anything proportional to $\tn _a$.)

\subsection{Finite space--time and the induced connection}

We extend the asymptotically constant spinor fields off $\scrif$ and into the finite space--time by requiring that they be transported parallel along the outgoing null congruence associated with the radiation field. 
The efficient way to implement this is to introduce a connection $D_a$ measuring the discrepancy from asymptotic constancy.  

We define $D_a\xi ^B$ be requiring $\zeta _BD_a\xi ^B=\nabla _a(\zeta _B\xi ^B)$ for every asymptotically constant spinor $\zeta ^B$.  That is, 
\begin{equation}
\zeta _BD_a\xi ^B=\zeta _B\nabla _a\xi ^B+\xi ^B\nabla _a\zeta _B\, .
\end{equation}
Evidently, our task is to compute $\nabla _a\zeta ^B$ for asymptotically constant spinors $\zeta ^B$.

The computation will be valid in the Bondi chart, and will use parallel propagation along the geodesic congruence with tangent field $l^a$.  For
any point $p$ in the chart, and any vector $u^a$ at $p$, let 
$u^a(q)$ be the connecting (Jacobi) field along the geodesic, so $l\cdot\nabla u^a=u\cdot\nabla l^a$.  We then have, since $\zeta ^B$ is covariantly constant along $l^a$, that
\begin{equation}\label{cueq}
  l^e\nabla _e( u^f\nabla_f \zeta ^B ) =l^eu^fR_{efQ}{}^B\zeta ^Q\, ,
\end{equation}
where $R_{efQ}{}^B=(1/2)R_{efQQ'}{}^{BQ'}$ is the curvature acting on spinors.  We may regard eq. (\ref{cueq}) as an evolution equation for $u\cdot \nabla\zeta ^B$ along the outward geodesic $\gamma$ through $p$.  Indeed, we have
\begin{equation}\label{icueq}
 u\cdot \nabla \zeta ^B\Bigr|^{\gamma (s_1)}_{\gamma (s_0)}
   =\int _{s_0}^{s_1} l^eu^fR_{efQ}{}^B\zeta ^Q\, ds\, ,
\end{equation}
where we understand that parallel transport along $\gamma$ is used to relate quantities at different points along this geodesic.

In order to put this in a more useful form, let us introduce a Green's function for the connecting fields:  let $W^a{}_b(q,p)$ be such that
$u^a(q)=W^a{}_b(q,p)u^b(p)$ is the connecting field along $\gamma$ which is $u^a(p)$ when $q=p$, that is
\begin{equation}\label{opt}
\left.\begin{array}{ccc}
  l\cdot\nabla W^a{}_b&=&(\nabla _cl^a)W^c{}_b\\
W^a{}_b(p,p)&=&\delta ^a{}_b
\end{array}\right\}\, ,
\end{equation}
where the operator $l\cdot\nabla$ acts on the variable $q$.  We note for future use that
\begin{equation}
\nabla _cl^a=-l_c\tau {\overline m}^a 
  +m_c(\rho {\overline m}^a+{\overline\sigma} m^a)
  +\,\mbox{conjugate}\, ,
\end{equation}
and so (\ref{opt}) can be integrated if the spin-coefficients are known.

We can now rewrite eq. (\ref{icueq}) as
\begin{eqnarray}
  \nabla _a\zeta ^B\Bigr| _{p=\gamma (s_0)}
 & =&W^c{}_a(q,p)\nabla _c\zeta ^B\Bigr| _{q=\gamma (s_1)}
   \\
 &&    -\int _{s_0}^{s_1} l^eW^f{}_a(\gamma (s),p)R_{efQ}{}^B(\gamma (s))\, ds\, \zeta ^B\, ,\nonumber
\end{eqnarray}
where again parallel transport along $\gamma$ is understood to compare the quantities at different points.  The idea is now to take $q$ to $\scrif$.  In this case, the first term on the right will be given by the results of the previous section, and we have
\begin{eqnarray}
\nabla _a\zeta ^B\Bigr| _{p=\gamma (s_0)} &=&
  -\lim _{s\to +\infty} W^c{}_a(\gamma (s),p) {\tilde m}_c{\dot{\overline\sigma}}^0\omicron _Q\omicron ^B\zeta ^Q\label{coneq}\\ 
&&   -\int _{s_0}^{+\infty} l^eW^f{}_a(\gamma (s),p)R_{efQ}{}^B(\gamma (s))\, ds\, \zeta ^Q\, .\nonumber
\end{eqnarray}
Notice that the right-hand side involves $\zeta ^Q$ only algebraically.

The formula (\ref{coneq}) is exact, and is valid in any Bondi chart.  In principle, it can be evaluated if the ``optical'' quantities $\rho$, $\sigma$ and $\tau$, governing the evolution of the pencil of null geodesics near $\gamma$, and the curvature, are known.  
Here we are only interested in its form in the radiation zone.

In this computation, we must keep track of the error terms $O(r^n)$ for both $r(p)$ and $r(q)$ in $W^a{}_b (q,p)$.  The easiest way to do this is to consider a basis $U_j{}^a(q)$ of connecting fields ($j=0,1,2,3$); then $W^a{}_b(q,p)=(U^{-1})_b{}^j(p)U_j{}^a(q)$.\footnote{While of course this formula is exact, the precise error estimates one gets depend on how well one knows the solutions $U_j{}^a$, which generally depends both on the basis chosen and how many terms in the asymptotic expansion one wants.  We are simply interested in getting the dominant contribution and ensuring that it is dominant, and there is a natural choice which is adequate for this.}  
From the standard expansions
\begin{eqnarray}
  \rho &=& -r^{-1} +O(r^{-3})\label{rhoest}\\
  \sigma &=&\sigma ^0 r^{-2}+O(r^{-3})\label{sigest}\\
  \tau &=&-(1/2)\Psi _1^0r^{-3}+O(r^{-4})\, .\label{tauest}
\end{eqnarray}
it is easy to check that we may take
\begin{equation}\label{uueq}
U_j{}^a=\left[\begin{array}{cccc}
  1&0&0&0\\
  0&1&O(r^{-2})&O(r^{-2})\\
  0&0&r+O(1)&O(1)\\
  0&0&O(1)&r+O(1)\end{array}\right]
  \left[\begin{array}{c}
    l^a\\ n^a\\ m^a\\ {\overline m}^a
    \end{array}\right]\, .
\end{equation}    
Then
\begin{eqnarray}\label{uieq}
(U^{-1})_b{}^j&=&
  \left[\begin{array}{cccc}
    n_b& l_b& -{\overline m}_b& -m_b
    \end{array}\right]\\
  &&\times \left[\begin{array}{cccc}
  1&0&0&0\\
  0&1&O(r^{-3})&O(r^{-3})\\
  0&0&r^{-1}+O(r^{-2})&O(r^{-2})\\
  0&0&O(r^{-2})&r^{-1}+O(r^{-2})\end{array}\right]\, ,\nonumber
\end{eqnarray}    
and 
\begin{eqnarray}\label{ffW}
W^a{}_b(q,p) &=&n_bl^a+l_bn^a-\frac{r(q)}{r(p)}\left( {\overline m}_bm^a+m_b{\overline m}^a\right)\nonumber\\
&&\qquad+\mbox{lower-order terms}\, ,
\end{eqnarray}
where the estimates on the lower-order terms can be recovered by multiplying (\ref{uueq}) and (\ref{uieq}), if needed.  One should bear in mind that on the right-hand side of eq. (\ref{ffW}), the vectors are evaluated at $q$ and the covectors at $p$.

We then have, for the first term in (\ref{coneq}), the expression
\begin{eqnarray}
-\lim _{s\to +\infty} W^c{}_a(\gamma (s),p) {\tilde m}_c{\dot{\overline\sigma}}^0\omicron _Q\omicron ^B
  &=&-\frac{{\dot{\overline\sigma}}^0}{r(p)}m_a\omicron _Q\omicron ^B\nonumber\\ 
&& +O(r(p)^{-2}).
\end{eqnarray}
We also have, using Sachs peeling, that
\begin{equation}
-\int _{r(p)}^\infty l^eW^f{}_a(q,p)R_{efQ}{}^B\, ds
  =- \frac{\Psi _3^0}{r(p)}l_a\omicron _Q\omicron ^B
    +O(r(p)^{-2})\, .
\end{equation} 
We recall that $\Psi _3^0=-\teth \dot{\overline\sigma} ^0$, and that the news function $N=-\dot{\overline\sigma}^0$.  Thus we have
\begin{equation}\label{aconstspin}
\nabla _a\zeta ^B=r^{-1}\left( Nm_a-\teth N l_a\right)\omicron _Q\omicron ^B \zeta ^Q +O(r^{-2})
\end{equation}
and 
\begin{equation}
  D_a\xi ^B =\nabla _a\xi ^B -r^{-1} \left( Nm_a-(\teth N)l_a\right)\omicron _Q\omicron ^B\xi ^Q+O(r^{-2})\, .
\end{equation}
Alternatively, using the standard far-field forms $\sigma '=N/r$, $\kappa ' =\teth N /r$ for the shear and acceleration of the field $n^a$, we can write
\begin{equation}\label{Dconeq}
  D_a\xi ^B =\nabla _a\xi ^B -\left( \sigma 'm_a-\kappa ' l_a\right)\omicron _Q\omicron ^B\xi ^Q +O(r^{-2})\, ,
\end{equation}
and so on vectors
\begin{eqnarray}\label{Dconeqvec}
D_a\xi ^b&=&\nabla _a\xi ^b
  -\bigl( (\sigma 'm_a-\kappa 'l_a)(l_qm^b-m_ql^b)\xi ^q
\nonumber\\
&&+\,\text{conjugate}\bigr) +O(r^{-2})\, .
\end{eqnarray}

Equation (\ref{Dconeqvec}) gives the parallelism in the radiation zone which enables us to compare local energy--momenta at different events.  This connection is by construction curvature-free and metric-preserving; it does (therefore) have torsion.

\subsection{Asymptotic frames}

Using the formula (\ref{Dconeq}) for the connection defining the asymptotically constant spinors and vectors, we have, in the radiation zone,
\begin{eqnarray}
  D_a\omicron ^B&=&\rho m_a\iota ^B -(\alpha m_a+\beta {\overline m}_a)\omicron ^B+O(r^{-2})\\
 D_a\iota ^B&=&\rho '{\overline m}_a\omicron ^B
  +(\alpha m_a+\beta {\overline m}_a)\iota ^B +O(r^{-2})
\end{eqnarray}
where all the spin-coefficients which appear have, to the required order, the same forms as they do in Minkowski space.
(This would not be true had we used the covariant derivative $\nabla _a$ in place of $D_a$).  Thus the asymptotically constant spinor or vector fields are given by linear combinations of the dyad or tetrad elements, where the coefficients are spherical harmonics (spin-weighted, in the spinor case).

We remark that in particular the Bondi--Sach frame's time-vector $t^a=(1/2)l^a+n^a$ is asymptotically constant.  However, it is not asymptotically covariantly constant; one has
\begin{equation}
  \nabla _at^b=-\kappa 'l_am^b+\mbox{conjugate}\, .
\end{equation}
The integral curves of $t^a$ are not geodesics; rather
\begin{equation}
 t^a\nabla _at^b=-\kappa ' m^b+\mbox{conjugate}\, .
\end{equation}   
Here and from now on, we drop the qualifier ``$+O(r^{-2})$.''

\section{Energy--momentum exchange}

We can now compute the exchange of energy--momentum between matter and the gravitational field in the radiation zone.  

Let $\zeta ^a$ be any asymptotically constant vector field, so $D_a\zeta ^b=0$.  Then if $T_{ab}$ is the stress--energy, the quantity $T_{ab}\zeta ^b$ can be interpreted as the four-current 
of 
local material
energy--momentum 
in the $\zeta ^a$ direction.  Thus $\nabla ^a(T_{ab}\zeta ^b)$ will give the rate of creation of the $\zeta ^a$-component of material energy--momentum per unit time per unit volume, and
because the total energy--momentum of the system (including gravitational radiation) is fixed, we attribute this creation to a conversion of gravitational energy--momentum.  We have, from eq. (\ref{aconstspin}),
\begin{eqnarray}
\nabla ^a(T_{ab}\zeta ^b)
  &=&T_{ab}\nabla ^a\zeta ^b\nonumber\\
  &=&T_{ab}(\sigma ' m^a-\kappa ' l^a)(l_dm^b-m_dl^b)\zeta ^d
    \nonumber\\ &&\quad+\mbox{conjugate}\, ,
\end{eqnarray} 
and thus
\begin{equation}\label{emomex}
\frac{d{\mathcal P}_d}{d\tau}=  T_{ab}(\sigma ' m^a-\kappa 'l^a)(l_dm^b-m_dl^b)
    +\mbox{conjugate} 
\end{equation}
is the rate of conversion of gravitational to material energy--momentum per unit time per unit volume (here $d\tau$ is the four-volume element).  (More precisely, it is the rate of conversion of gravitational, to material, contributions to the total energy--momentum, but this is too  cumbersome to say each time.)
In particular, the rate of conversion of gravitational to material energy (with respect to the Bondi frame, the $t^d=(1/2)l^d+n^d$ component of (\ref{emomex})) per unit time per unit volume is
\begin{equation}\label{eex}
\frac{d{\mathcal E}}{d\tau}=  T_{ab}(\sigma ' m^a-\kappa 'l^a)m^b
    +\mbox{conjugate}\, . 
\end{equation}
It should be emphasized that this does not mean that matter is created or destroyed; it rather means that the contribution of whatever matter is present to the energy--momentum of the system, as measured at null infinity, may change.  

For matter confined to the region under study, we may compute the total rate of conversion $d{\mathcal P}_a/du$ of gravitational to material energy--momentum per unit retarded time:
\begin{eqnarray}
\frac{d{\mathcal P}_d}{du} \zeta ^d&=&\int _{u=\const} T_{ab}(\sigma ' m^a-\kappa 'l^a)(l_dm^b-m_dl^b)\zeta ^d\nonumber\\ 
&&\quad +\,\mbox{conjugate}\, ,\label{totemis}
\end{eqnarray}
where $\zeta^a$ is any asymptotically constant vector and the volume form $\epsilon _{abcd}n^a$ is understood.  

The fact that only the vectors $l^a$, $m^a$, ${\overline m}^a$ appear in these expressions has implications for both which components of the stress--energy contribute to the energy--momentum exchange, and how the local contributions are directed.
Matter moving ultrarelativistically outward from the source (which will have $T_{ab}$ proportional to $l_al_b$) will not exchange energy--momentum with gravitational radiation;\footnote{This may be contrasted with the interconversion of electromagnetic and gravitational waves in an electromagnetic background, e.g. ref. \cite{GP1980}.} inward-directed 
ultrarelativistic matter or radiation (with $T_{ab}$ proportional to $n_an_b$) will tend to exchange transverse momentum (in the $m^a$--${\overline m}^a$ plane); transverse stress--energy components are needed to exchange outward-directed momentum.  
Depending on the sign of the energy--momentum exchange, the effect on a local distribution of matter may be to increase its energy and contribute to its momentum outward, or to decrease its energy and contribute to its momentum inward:  but it cannot increase the energy and also contribute to inward momentum, nor decrease the energy and contribute to outward momentum.

The rate of energy--momentum exchange thus is given by a sum of terms, each of which is a product of a component of the stress--energy with the news function (or an angular derivative of the news function).  It is thus {\em formally first-order} in both the matter and the gravitational radiation.  On the other hand, one must remember that the entire analysis assumes that the curvature effects of the matter terms only give small perturbations to the geometry set by the Bondi--Sachs asymptotics,
and thus the exchange has only been established in cases where it is {\em effectively second-order} in the radiation.

There are some further fine points about the order of the effects involved, which are discussed next.

\section{Relation to linearized theory}

Since in most practical cases gravitational radiation is weak, it is natural to analyze it by perturbation theory.  This can certainly be done within the framework developed here.  This is for the most part straightforward, but there are two points which deserve comment.

The first point is elementary but worth making explicitly:  the holonomies in the asymptotic regime are non-trivial at the linearized level.  This means that the construction of the asymptotically constant covectors will also be non-trivial at this level, as will be energy--momentum-exchange effects.  In other words, the essential physics is non-local, but it is not exclusively non-linear.

The second point is  
a conceptual
one which will not figure explicitly in the analysis but explains the perhaps unexpected forms of some of the results.
The reader may wish to skim this at first and come back to it as necessary.  

In the Bondi--Sachs analysis, the power radiated is given by
\begin{equation}
  (4\pi G)^{-1}\lim _{r\to\infty}\oint |\sigma '|^2 \, d{\mathcal S}
\end{equation}
over spheres $r=\const$, $u=\const$.
(There are similar formulas for the other components of the radiated energy--momentum.)
Because this is quadratic in $\sigma '$, one might think that a knowledge of $\sigma '$ in linearized theory would be adequate for a lowest-order computation of the power.  In particular, it would be tempting to think of the physics in the radiation zone as due to two contributions, one from a central radiating source and the other from the matter in the zone.  We are already assuming the matter effects are smaller than the radiation ones; if (as will usually be the case) the radiation from the central source is also small, shouldn't we have simply $\sigma '=\sigma '_{\rm central\ source}+\sigma '_{\rm radiation-zone\ matter}$?  And wouldn't energy--momentum exchange effects be derivable from such considerations?

The answer is No.  To understand this, suppose we start with a vacuum radiation zone, and ask how the outgoing shear $\sigma '$ there changes as a little matter is introduced.  To do this, we follow the geometry from outside the matter inwards, along the $l^a$ congruence.  As we do so, we must maintain the Bondi coordinate and tetrad conditions.  In particular, the tetrad vectors $n^a$ and ${\overline m}^a$, which figure in the definition of $\sigma '$, will be affected by the matter.  Thus there will be changes in what we would call $\sigma '_{\rm central\ source}$ due to the change in the tetrad vectors as they are affected by the matter's gravitation.  These will lead to energy--momentum exchanges of the same order as, but distinct from, the cross-terms $\sigma '_{\rm central\ source}{\overline\sigma}'_{\rm radiation-zone\ matter}$ (plus conjugate).  (There will generally be other changes, too, of the same magnitude.)

In the following sections, we will have examples of this.

\section{Test particles}

Test particles can be viewed as special cases of the general results of the previous section, or analyzed directly in terms of the asymptotically constant frame and the connection $D_a$.

\subsection{General formulas and interpretations}

Let a particle of mass $\mu$ fall freely along an 
affinely parameterized geodesic $\gamma$.  Then the geodesic equation 
implies  $\dot\gamma ^b\nabla _b (\mu\dot\gamma _a)=0$, which means 
that the energy--momentum of the particle is propagated parallel along 
the trajectory.  On the other hand, the rate of change of the 
energy--momentum {\em with respect to the Bondi--Sachs frame} is 
\begin{eqnarray}\label{momenteq}
  \dot\gamma ^bD_b(\mu\dot\gamma _a)&=&\mu\left(\kappa ' l_b-\sigma 'm_b\right)\left( l_am^c-
m_al^c\right)
\dot\gamma ^b\dot\gamma _c\nonumber\\
&&\quad +\,\text{conjugate}\, . 
\end{eqnarray}

To begin to understand this, it may be helpful to compare it with the formula for the motion of a particle of charge $q$ in a Minkowski-space electromagnetic radiation zone, which is
\begin{equation}
  \dot\gamma ^b\nabla _b (\mu\dot\gamma _1)
=q\phi _2 (l_am^c-m_al^c){\dot\gamma}_c+\,\text{conjugate}\, ,
\end{equation}
where $\phi _2$ is the radiative component.  Thus the two cases have a common factor $(l_am^c-m_al^c)\dot\gamma _c$;
this codes the polarization in the electromagnetic case, and is essentially the square root of the polarization in the gravitational one.  The remaining directional character of the gravitational effects appears in the factor $\kappa'l_b-\sigma 'm_b$.

In the Maxwell case it is a component of the Faraday tensor,
a local geometric object, which enters: but for gravity it 
is the spin-coefficients $\sigma '$ and $\kappa '$, which are essentially nonlocally determined potentials for the curvature, which come up.  In particular, the Bondi news, and so $\sigma '$, may be non-zero in a range of $(u,\theta ,\phi )$ values for which the radiative curvature term $\Psi _4^0$ vanishes.  

In the non-relativistic limit (where $\dot\gamma ^a$ differs from $t^a$ only by small terms), one has to leading order
\begin{equation}
\dot\gamma ^bD_b (\mu\dot\gamma _a)
=-\mu\kappa 'm_a+\,\text{conjugate}\, ,
\end{equation}
which is formally similar to the response
$-q\phi _2m_a+\,\text{conjugate}$ of a charged particle; however, for relativistic motion the difference in forms between the equations becomes apparent.
For ultrarelativistic particles, the geometric factors $m\cdot\dot\gamma$, $l\cdot\dot\gamma$ will tend to zero along the outgoing portions of the trajectory, and these will inhibit energy--momentum exchange. 

In eq. (\ref{momenteq}), the differentiation is with respect to the affine 
parameter (say $s$) along the geodesic.  If we convert it to 
differentiation with respect to the Bondi parameter $u$ using $\rmd u/
\rmd s=\dot\gamma ^a l_a$, we have
\begin{eqnarray}
{D\over{Du}}\mu\dot\gamma _a
&=&-\sigma '\left(\mu\frac{(\dot\gamma ^bm_b)^2}{\dot\gamma ^c l_c} 
\right)
  l_a +\kappa '(\mu\dot\gamma ^bm_b)l_a\nonumber\\
  &&+\mu\dot\gamma ^b \left( \sigma ' m_am_b-\kappa 'm_al_b\right)
  \nonumber\\
 &&
  +\,\text{conjugate}\, .\quad\label{emex}
\end{eqnarray} 
This expression represents the rate of transfer of energy--momentum 
from
the gravitational field to the particle.  (While the particle's
energy--momentum changes relative to the Bondi--Sachs frame, its mass
$\sqrt{\mu {\dot\gamma}_a\mu{\dot\gamma}_b g^{ab}}$
does not, because the connection $D_a$ preserves the metric.)

The terms on the right in~(\ref{emex}) have interesting interpretations.  The 
factor
$\mu (\dot\gamma ^bm_b)^2/\dot\gamma ^cl_c$ in the first term would 
be,
in linearized theory, half the contribution of the test particle to the 
Bondi
shear.  Thus there is a shear--shear coupling, between the shear $
\sigma
'$ of the {\em ingoing} congruence and a measure of the Bondi (outgoing) shear due to 
the
test particle, leading to a transfer of energy--momentum along $l_a$.
The second term gives also an outward-directed acceleration (with respect to the Bondi frame), this one proportional to the acceleration $\kappa '$ of the ingoing congruence.  The terms in parentheses of the second line of eq. (\ref{emex}) are half the projection of ${\mathcal L}_ng_{ab}$, the Lie derivative of the metric along $n^a$, in the directions spanned by $m_am_b$ and $m_al_b$.  Roughly speaking, the effect of the second line is as if the particle experienced an acceleration (relative to the Bondi frame) from being batted by the {\em ingoing} congruence, or rather by the projected effects of this.
(One should bear in mind that the incoming congruence does not here code any incoming radiation, but rather the temporal evolution which is necessary to maintain the Bondi gauge.)

Over short portions of the particle's trajectory, we expect the
geometric terms $\dot\gamma ^am_a$, $\dot\gamma ^al_a$ to be nearly
constant, as well as the value of $r$.  We have $\sigma '\simeq
-{\dot{\overline \sigma}}_{\rm B}/r$
and $\kappa '\simeq -\teth {\dot{\overline\sigma}}_{\rm B}/r$
in the radiation zone.  Over 
short
portions of the trajectory, the angular variables which (with $u$) are
the arguments of $\sigma _{\rm B}$ do not change much, and so the 
change
in the particle's energy--momentum as measured with the Bondi--Sachs
frame will be
\begin{eqnarray}
\Delta \mu\dot\gamma _a&\approx&
\frac{\Delta{\overline\sigma}_{\rm B}}{r}\left( \mu
  \frac{({\dot\gamma}^bm_b)^2}{\dot\gamma ^cl_c}\right) l_a
-\frac{\teth\Delta{\overline\sigma}_{\rm B}}{r}
  (\mu\dot\gamma ^b m_b)l_a\nonumber\\
&&+\mu\dot\gamma ^b\left(
 -\frac{\Delta{\overline\sigma}_{\rm B}}{r} m_am_b
  +\frac{\teth\Delta{\overline\sigma}_{\rm B}}{r} m_al_b
  \right)\nonumber\\
&&+\,\text{conjugate}\, .
\end{eqnarray}
This means that we have an approximately conserved quantity
\begin{eqnarray}
\Pi _a&=&\mu\dot\gamma _a
+
\frac{{\overline\sigma}_{\rm B}}{r}\left( \mu
  \frac{({\dot\gamma}^bm_b)^2}{\dot\gamma ^cl_c}\right) l_a
-\frac{\teth{\overline\sigma}_{\rm B}}{r}
  (\mu\dot\gamma ^b m_b)l_a\nonumber\\
&&+\mu\dot\gamma ^b\left(
 -\frac{{\overline\sigma}_{\rm B}}{r} m_am_b
  +\frac{\teth{\overline\sigma}_{\rm B}}{r} m_al_b
  \right)\nonumber\\
&&+\,\text{conjugate}\, .
\end{eqnarray}

The existence of this approximate conservation law is closely 
connected
with the character of the back-reaction of the test particle on the
radiation field.  Were the quantity to be exactly conserved, one would
expect no radiated energy, and hence no back-reaction, in cases where $\sigma _{\rm B}$ returns to its original value after a burst of radiation.  A net change in $\sigma _{\rm B}$ would be a ``memory effect''; we see that glitches leading to steps in $\sigma _{\rm B}$ over short $u$-intervals (over which $\Pi _a$ is conserved) correspond to steps in the particle's energy--momentum, measured relative to the Bondi--Sachs frame.

\subsection{Scattering by quadrupolar waves}\label{quadrupsec}

A simple but important example is the scattering of test particles by quadrupole radiation in linearized gravity.

Because the energy--momentum transfer is linear in the news, it will suffice to consider the case of a constant polarization; the general case is a sum of such terms.  
We take the Bondi news to be $N=f(u)K^{cd}{\overline m}_c{\overline m}_d$, where the amplitude profile $f(u)$ is a complex dimensionless function, and the polarization $K^{cd}$ (also dimensionless) is
a fixed Minkowskian tensor, real, symmetric, trace-free 
and
orthogonal to $\partial /\partial t$; here $u=t-r$ is the retarded time.
The real and imaginary parts of $f$ determine, respectively, what are called the {\em electric} and {\em magnetic} contributions to the news.
We have
\begin{eqnarray}
\sigma '&=& \frac{f(u)}{r} K^{cd}
  {\overline{m}}_c
  {\overline{m}}_d\\
 \kappa '&=&-2\frac{f(u)}{r} K^{cd}
  {\overline{m}}_c  {\hat r}_d
\, .
\end{eqnarray}

Since the energy--momentum transfer is proportional to the news function, which is already first-order, we may take the particle's trajectory to be Minkowskian.
Let us write 
\begin{equation}
\gamma ^a (s) =bB^a +s(\cosh\xi )t^a
  +s(\sinh\xi )C^a
\end{equation}
as a vector relative to the origin in Minkowski space, where $B^a$,
$C^a$ are mutually orthogonal spacelike unit vectors, orthogonal
to $t^a=\partial /\partial t$, the impact
parameter is $b$ and the rapidity is $\xi$.  
We have then
\begin{eqnarray}
 r&=&\sqrt{b^2+s^2\sinh ^2\xi}\\
  u&=&s\cosh\xi -\sqrt{b^2+s^2\sinh ^2\xi }\, .
\end{eqnarray}
We also note for future use that
\begin{equation}
{\hat r}^a=(1/2)l^a-n^a
 =\left(bB^a+s(\sinh\xi )C^a\right)/r
\end{equation}
is a unit radial vector orthogonal to $t^a$, and then
\begin{eqnarray}\label{mident}
{\overline m}_cm_a &=&
  {\overline m}_{(c}m_{a)} +{\overline m}_{[c}m_{a]}\nonumber\\
  &=& (1/2)\left( -g_{ca} +t_ct_a- {\hat r}_c{\hat r}_a 
   -\rmi \epsilon _{capq}t^p {\hat r}^q\right) \, .\quad
\end{eqnarray}

The energy--momentum transfer is then
\begin{eqnarray}
\Delta P_a&=&\mu\int [\kappa ' l_p-\sigma 'm_p][m_ql_a- l_q m_a]
{\dot\gamma}^p{\dot\gamma}^q
\, ds\nonumber\\ 
&&\quad +\mbox{conjugate}\nonumber\\
&=&-2\mu\int\frac{f(u)}{r}K^{cd}{\dot\gamma}^p{\dot\gamma}^q
{\overline m}_cm_{[q}l_{a]}({\overline m}_dm_p+2{\hat r}_dl_p)\, ds\nonumber\\ 
&&\quad +\mbox{conjugate}\label{qtr}
\, .
\end{eqnarray}
Note that it is orthogonal to $P^a$ (as expected, since we consider only first-order changes and $P_aP^a$ is preserved).

For a given radiation field, the dependence of the energy--momentum transfer on the trajectory $\gamma$ is quite rich, and we shall here work out only a few limiting cases.

{\em The limit $\xi \to 0$.}  This is the non-relativistic limit touched on earlier.
In this case we have ${\hat r}^a=B^a$, $r=b$, $ds=du$, as well as
\begin{eqnarray}
 {\dot\gamma}^q{\overline m}_cm_{[q}l_{a]}
   &=&-(1/4)\times\\
   &&(-g_{ca}+t_ct_a-B_cB_a-i\epsilon _{cars}t^rB^s)\nonumber\\
 {\dot\gamma}^p({\overline m}_dm_p+2{\hat r}_dl_p)
 &=&2B_d\, ,
\end{eqnarray}
so
\begin{eqnarray}\label{stattrans}
\Delta P_a&=&(\mu /b)K^{cd}  (-g_{ca}-B_cB_a-i\epsilon _{cars}t^rB^s) B_d\times\nonumber\\
&&\quad\int f(u)\, du +\mbox{conjugate}\, .
\end{eqnarray} 

As noted earlier, the momentum transfer in this case is a memory effect, responding to
$\Delta\sigma ^0 =-K^{cd}m_cm_d\int {\overline f} du$, the net change in Bondi shear.
The transfer is orthogonal to $B_a$, and falls off with $b$, the distance from the source.
The real and imaginary parts of the news couple through the real and imaginary terms within the parentheses in (\ref{stattrans}),
giving different parity-dependences on the separation vector $B^a$.

{\em The case $|\xi |\to\infty$.}  This case is much more complicated.  The calculations are straightforward, though, and I shall only indicate the main points.

The energy--momentum transfer is
\begin{eqnarray}\label{ultra}
\Delta P_a&=&-2\mu e^{|\xi|}\int _{-\infty}^0 f(u) K^{cd}(A_{ca}+B_{ca})D_d\frac{du}{|u|}\nonumber\\
&&\quad+\mbox{conjugate}
\end{eqnarray}
(one can check that the numerator in the integrand vanishes as $u\uparrow 0$, and in fact there is no singularity at $u=0$),
where
\begin{eqnarray}
A_{ca}&=&(1/4)\sign (\xi )\sech\zeta \nonumber\\
&&
\times (-\sech\zeta C_c+\sign (\xi )\tanh\zeta B_c-i\epsilon _{cqrs}C^qt^rB^s)\nonumber\\
&&\times
(t_a+\sech \zeta B_a+\sign (\xi )\tanh\zeta C_a)\\
B_{ca}&=&(1/4)(-g_{ca}+t_ct_a-{\hat r}_c{\hat r}_a
  -i\epsilon _{capq}t^p{\hat r}^q)\nonumber\\
&&\times (1-\tanh\zeta )\\
D_d&=&(1/4)\sign (\xi )\sech\zeta \nonumber\\
&&\times (-\sech\zeta C_d+\sign (\xi )\tanh\zeta B_d-i\epsilon _{dpqr}C^pt^qB^r)\nonumber\\
&&+(1-\tanh\zeta)(\sech\zeta B_d+\sign (\xi )\tanh\zeta C_d)
\end{eqnarray}
with
\begin{equation}
 \zeta =-\log (|u|/b)\, .
\end{equation}
Note that because this scales as $e^{|\xi|}$, the {\em fractional} energy-transfers $\Delta E/E$ will attain a $|\xi|$-independent limit (and this will also apply to massless particles).  

While the general form of the energy--momentum transfer is evidently complicated, it simplifies considerably in certain regimes.  For any fixed trajectory, the contributions to $\Delta P_a$ from different values of $u$ break down to those from an {\em incoming regime} $u\lesssim -b$, a {\em transition regime} $u\sim -b$, and an {\em outgoing regime} $-b\lesssim u <0$.  It is only in the transition regime that all the terms in the integrand are potentially significant; in the other two the limiting forms are much simpler.
I will give the forms first, and then discuss their interpretations.

The contribution from the incoming regime is
\begin{eqnarray}\label{ultrao}
\Delta P_a\Bigr|_{\rm incoming}&=&-2\mu e^{|\xi |}K^{cd} \nonumber\\
&&\times(-g_{ca}-C_cC_a-i\sign (\xi )\epsilon _{capq}t^pC^q)\nonumber\\
&&\times(\sign (\xi ) C_d)\int_{u\lesssim -b}f(u)\frac{du}{u}
  \nonumber\\
&&\quad +\mbox{conjugate}\, ,\quad
\end{eqnarray}
and from the outgoing regime
\begin{equation}\label{ultraa}
\Delta P_a\Bigr|_{\rm outgoing}=\zz P_a
\end{equation}
where
\begin{equation}\label{ultrab}
\zz = 
(2b^2)^{-1}K^{cd}U_cU_d\int _{-b\lesssim u<0} f(u)u\, du 
+\mbox{conjugate}
\end{equation}
with
\begin{equation}\label{ultrac}
U_c=B_c-i\sign (\xi )\epsilon _{cqrs}C^qt^rB^s\, .
\end{equation}
The notation $\zz$ is chosen to fit with the usual notion of a red-shift.  
Thus formula (\ref{ultrao}) would suffice to give the energy--momentum transfer if the amplitude $f$ vanished for $-b\lesssim u <0$, and (\ref{ultraa}) would suffice if the amplitude vanished for $u\lesssim -b$.

The contributions from the incoming regime have the following important features:  (a) They are independent of the direction $B^a$ of the trajectory's closest approach.  
(b) The directional dependence is formally the same as in the $\xi =0$ case, but with $B^a$ there replaced by $\sign (\xi )C^a$ here.  In particular, the transfer is purely one of spatial momenta.
(c) They depend on the impact parameter $b$ only through the range of integration. 

The contributions from the outgoing regime are very different:  (a) They purely dilate the energy--momentum, that is, they purely red- or blue-shift it, without changing its space--time direction.  (b) Their angular dependence is very curious, with the electric part coupling to $B_cB_d-W_cW_d$ (where $W_c=\epsilon _{cqrs}C^qt^rB^s$; note that $W^c =(B\times C)^c$ in three-vector terms), and the magnetic part to $\sign (\xi )(B_cW_d+B_dW_c)$. 

One would like to get a sense of what the range of the scattering is, in terms of how it depends on the impact parameter.  Because the source will be time-dependent, there is no truly universal answer to this.  If we consider a fixed source and formally expand the exchange (\ref{ultra}) for $b\to\infty$, we find it scales as
$\int _{-\infty}^0 f(u)udu /b^2$ for $b\to\infty$.  If for instance $f$ were compactly supported, then this would show that the scattering fell off as $1/b^2$.  This is more rapid than the Newtonian result, because the Newtonian scattering accumulates over a large portion of the particle's trajectory, with significant contributions over a spatial scale $\sim b$.  (If the Newtonian force somehow acted only for a time-interval $\Delta t$ near the particle's point of closest approach, the scattering would go as $v\Delta t/b^2$, with $v$ the particle's speed.)  For a monochromatic source, because the scattering will average out for the early portion of the particle's trajectory, a similar argument applies, and we expect a scaling $\sim 1/(\omega b)^2$ for angular frequency $\omega$.  However, a wave which falls off as $|f(u)|\sim |u/u_0|^{-1/2 -\epsilon}$ as $u\to -{\infty}$ will carry finite energy but could by eq. (\ref{ultrao}) lead to exchanges scaling as $|u_0/b|^{-1/2-\epsilon '}$.

\section{Nonzero-average effects}

We have seen that the energy--momentum exchange between matter and gravitational waves is determined by integrals of components of the stress--energy against certain spin-coefficients, which are in turn proportional to the Bondi news $N$ (and its angular derivative $\teth N$).  If the frequencies of the gravitational waves are large compared to the scales on which the stress--energy changes, averaging will suppress the exchange.  
Interesting net exchange effects therefore require defeating this averaging. 

One possibility for doing this is to have matter which is not static on the time-scale over which the waves cycle.  Most simply, the stress--energy may beat with or against monochromatic waves, leading to a secular exchange (compare \cite{Gertsenshtein1962,CMR1997}).  Of course, such effects require a resonant tuning which will not be generic.

Even for waves which are not monochromatic, one could have matter whose time-dependence was correlated with that of the wave in such a way as to give non-trivial net effects.  Such a possibility was suggested some time ago, in the case of tidal energy--momentum exchanges, by Press \cite{Press1979}, with the idea of mimicking the mirrors, wave-guides, etc., available for electromagnetic radiation.  Of course, given the basic assumption of this paper that matter perturbs the radiative geometry only slightly, we cannot expect here to find anything like the efficiency needed to construct a gravitational mirror or wave-guides; also, the use of bulk rather than tidal effects makes the analogy with electromagnetism more distant, and there is a difference in that the present techniques speak most directly to energy--momentum, not wave-forms.
Nevertheless,
we will see that it is in principle possible in at least some cases to have matter respond to gravitational waves so that the flow of energy--momentum is coherently modified, for example, to cause a net absorption of one component.  

Finally, and potentially most broadly,
one could get net changes in energy--momentum exchange if the waves carry ``memory.''  The memory effect in this case is a net change in Bondi shear between  the period prior to the wave and the one after it.  
In the quadrupole approximation, the Bondi shear is essentially the second time-derivative of the quadropole, so
systems emitting jets will generate gravitational waves with memory.

I shall give examples of these different non-zero average effects here, simply to give a sense of some of the features and issues which come up.  (As will become apparent, there are so many degrees of freedom that full treatments would be lengthy.)  We shall see in particular that the gravitational waves from relativistic jets may be affected by their propagation through matter.

\subsection{Beating with or against the waves}

The simplest examples of secular energy--momentum effects are constructed from quadrupoles in linearized gravity.  We suppose that we have a central quadrupole source at the spatial origin, and at coordinate value $r$ another, smaller, quadrupole.  Any orbital motion of this smaller quadrupole will be neglected here, since we are interested in gravitational waves of much higher frequencies than any orbital frequency (compare \cite{CMR1997}).  

The key relation is eq. (\ref{emomex}), 
 repeated here:
\begin{equation}
\frac{d{\mathcal P}_d}{d\tau}
=T_{ab}(\sigma ' m^a-\kappa 'l^a)(l_dm^b-m_dl^b)
 +\,\text{conjugate}\, .
\end{equation}
The idea will be to integrate this over a small spatial volume containing the smaller quadrupole.

The stress--energy terms will be approximated as localized at the small object.  
While in reality the object will have a finite size (large enough, in particular, that it is nowhere near its Schwarzschild radius), most of the details of its internal structure will be irrelevant, and we will as usual suppose that we may represent it by a spatial multipole distribution insofar as integrals of smooth quantities, not varying much on the scale of the object, against its stress--energy go.
We will discard any monopole or dipole terms, since these are not expected to change rapidly enough to beat against the waves.
We will assume the remaining terms can be treated as pure quadrupoles.
In general such terms in the stress--energy involve certain coefficient functions times spatial delta-functions, or times derivatives of spatial delta functions \cite{ADH2013}.  Here any derivatives of spatial delta functions will be discarded, since these will be integrated against the (spatially) slowly-varying $\sigma '$, $\kappa '$.  We may then write
\begin{eqnarray}
T_{ab}l^al^b&=&(1/2){\ddot Q}_{ab}^{\rm el}l^al^b\delta _{\rm sp}\\
  T_{ab}l^am^b&=&(1/2){\ddot Q}^{\rm el}_{ab}l^am^b\delta _{\rm sp}\\
  T_{ab}m^am^b&=&(1/2){\ddot Q}^{\rm el}_{ab}m^am^b\delta _{\rm sp}\, ,
\end{eqnarray}
where $\delta _{\rm sp}$ is a spatial delta function at the object's location and $Q^{\rm el}_{ab}(u)$ is the object's ``electric'' or ``mass'' quadrupole, and the dots are derivatives with respect to $u$.  (The ``magnetic'' or ``current'' quadrupole does not appear because it contributes only terms with derivatives of $\delta _{\rm sp}$.)  Note that in these equations, because the quadrupoles are purely spatial, we may replace $l^a$ with ${\hat r}^a$.

Now let us turn to the central source.  We will allow it to have a complex quadrupole moment 
\begin{equation}
\Q _{ab}=\Q^{\rm el}_{ab}+i\Q ^{\rm mag}_{ab}\, ,
\end{equation}
where $\Q ^{\rm el}_{ab}$, $\Q ^{\rm mag}_{ab}$ are purely spatial, symmetric, trace-free tensors.  
We have
\begin{eqnarray}
N&=&-\Q _{ab}^{(3)}{\overline m}^a{\overline m}^b\\
\teth N&=&+2\Q _{ab}^{(3)}{\hat r}^a{\overline m}^b
\, ,
\end{eqnarray}
where the superscript indicates the third retarded time-derivative.  

As a final preparatory step, we have from eq. (\ref{mident}) the identity
\begin{equation}
  m_a{\overline m}_p=(1/2)(-\Pi _{ap}-i\RR _{pa})\, ,
\end{equation}
where 
\begin{equation}
\Pi ^a{}_{b}=-m^a{\overline m}_m-{\overline m}^{a}m_b
\end{equation} 
is the projection to the transverse spatial directions and
\begin{equation}
\RR ^p{}_a=\epsilon ^p{}_{aqs} t^q{\hat r}^s
\end{equation}
is the generator of rotations in this plane, about ${\hat r}^a$ (with the usual orientation).

Integrating eq. (\ref{emomex}), we have, after some algebra
\begin{eqnarray}\label{DeltaPint}
\Delta {\mathcal P}_d
&=&-\frac{1}{4}\int {\ddot Q}^{\rm el}_{ab}\Q ^{(3)\rm el}_{pq}
  [(\Pi ^{ap}\Pi ^{bq}-\RR ^{pa}\RR ^{qb}
  -4{\hat r}^a{\hat r}^p\Pi ^{bq}) l_d\nonumber\\
&& \qquad -l^b(\Pi ^{ap}\Pi_d{}^q -\RR ^{pa}\RR ^q{}_d)] r^{-1}du
  \nonumber\\
&&  +\frac{1}{4}\int {\ddot Q}^{\rm el}_{ab}\Q ^{(3)\rm mag}_{pq}
  [(2\Pi^{ap}\RR ^{qb}-4{\hat r}^a{\hat r}^p\RR ^{qb})l_d
  \nonumber\\
&&  \qquad -l^b(\Pi ^{ap}\RR ^q{}_d+\RR ^{pa}\Pi ^q{}_d)] r^{-1}du
 \, .
\end{eqnarray}
Evidently the dependence on the polarizations, indicated by the terms in square brackets, can be rather complicated.  However, the initial ${\ddot Q}^{\rm el}_{ab}\Q ^{(3)}_{pq}$ factors show clearly the possibilities for constructive or destructive interference, if the frequencies are matched.

At resonance at an angular frequency $\omega$, ignoring the polarizations, the energy--momentum exchange scales as $\sim \omega ^5 Q^{\rm el}_{ab}\Q _{pq}\Delta u /r$.  This should be contrasted with the radiated energy of the central source, which goes as $\sim \omega ^6\Q _{ab}\Q _{pq}\Delta u$.  Thus the relative change is suppressed by two factors:  the (assumed) intrinsic weakness of the waves from the small object relative to those from the source; and $c/(\omega r)$, which will be small in the radiation zone.  While in many practical cases this relative change will certainly be tiny, there is no inherent reason for it to be so in all cases.  Also 
many of these small objects could, in principle, surround the source.

The effects of the polarizations are curious and different from those of tidal effects.  Because of the complications of the expressions, I will just discuss a few of the possibilities.

Suppose that the source quadrupole has principal axes along the coordinates, with degenerate eigenvalues in the $x$--$y$ plane.  
Then it turns out that if the small object is on the $z$-axis, there is no effect.  However, for an object along (say) the $x$-axis, the coupling to $\Q ^{(3)\,\rm el}_{ab}$ would be proportional to
\begin{equation}
({\ddot Q}_{xx}-{\ddot Q}_{yy})l_d
  -{\ddot Q}_{xz}{\hat x}_d+{\ddot Q}_{xy}{\hat y}_d\, .
\end{equation}
Thus if the object's quadrupole were purely  in the ``plus'' polarization relative to the $x$- and $y$-axes, it would give net changes in the energy--momentum proportional to $l^a=t^a+{\hat x}^a$.\footnote{Note that this orientation is {\em not} transverse to the waves from the central source --- those would be described by polarizations transverse to the $x$-axis.}  Thus the system could acquire or loose energy--momentum in this direction, depending on the phase matching.  It is not clear, in general, how this would be distributed among the system's components.  However,
one could imagine the central source and the small object joined by some framework, in which case presumably the entire system would move in response. (One needs the usual caveat here about relativistic systems not being strictly rigid.)
If one had two small objects, oppositely placed about the central source, then depending on their phases, one could have energy--momentum transfers purely in the $t^a$ or ${\hat x}^a$ directions.  

I have so far emphasized the case of resonant coupling, but it is possible to generalize this, because the formula (\ref{DeltaPint}) expresses the coupling of the wave from the central source and the small object in the time-domain.  This equation shows that if the time-dependences of the waves from the central source and the matter in the radiation zone are suitably correlated, energy--momentum exchange effects can build up.

It is worth noting that formula (\ref{DeltaPint}) and its consequences are {\em not} what one would get by using linearized-gravity computations of the news and the Bondi--Sachs energy--momentum-loss formula; this is an example of the second point raised in Section VI.

\subsection{Dynamically active matter}

It is natural to ask whether there are gravitational analogs of the familiar materials which control and modify the propagation of electromagnetic waves, that is, of {\em optically active} materials.  In the context of the current work, we cannot expect any exact analogy, for we study here not the changes of the waves themselves but of their energy--momentum.  We may say that matter is {\em dynamically active} if it affects the energy--momentum of gravitational waves passing through (or near) it.
Of course this term --- like its optical counterpart --- is so broad that strictly speaking all matter has this property.  We are really interested in knowing what the character of the activity is.

The results of the previous subsection show that there are at least some similarities with optical activity:  that in some circumstances matter may coherently alter the flow of energy--momentum in the waves.  However, there is a very substantial difference.

The energy--momentum exchanges are given in terms of the spin-coefficients $\sigma '$, $\kappa '$, which are not locally determined, because they depend on the Bondi tetrad (in particular on the spinor $\iota _A$).  To the extent this nonlocality is essential, matter in the radiation zone cannot causally adapt to radiation in its vicinity.  That is, while matter might happen to be positioned in the radiation zone of a specific source so as to effect particular changes in the flow of energy--momentum, one cannot contrive a distribution of matter guaranteed to respond in a prescribed way to arbitrary sources.

It is possible to partially compensate for this limitation, by using the relations (valid asymptotically)
\begin{equation}
  \partial _u\sigma ' =\Psi _4\, ,
  \quad 
  r\eth\Psi _4=\partial _u\kappa '\, .
\end{equation}
These relate the spin-coefficients to components of the curvature tensor.  Now, while the curvature tensor itself is locally determined, one must be cautious that one is here taking components again with respect to the Bondi tetrad.  However, in the radiation zone, the Weyl curvature is to good approximation
$\Psi _{ABCD}=\Psi _4\omicron_A\omicron _B\omicron _C\omicron _D$, so it is enough to know the outgoing spinor $\omicron _A$ to know $\Psi _4$, and this will be the case if we assume we know the outgoing wave-fronts, or equivalently the coordinate $u$, locally.  (The phase of $\omicron _A$ will not matter.)  However, the operator $\eth$ will not be well-determined locally, being subject to an ambiguity of addition of $O(1)$ multiples of $l\cdot\nabla$.

Again assuming the relevant matter terms can be taken to be ``electric'' quadrupoles, we have
\begin{eqnarray}\label{partint}
\Delta {\mathcal P}_d&=&
(1/2)\int _{u_0}^{u_1} {\ddot Q}_{ab}^{\rm el}
(\sigma ' m^a-\kappa 'l^a) (l_dm^b-m_dl^b)\, du\nonumber\\
&&\quad+\,\text{conjugate}\nonumber\\
&=&(1/2){\dot Q}_{ab}^{\rm el}
(\sigma ' m^a-\kappa 'l^a) (l_dm^b-m_dl^b)\Bigr| _{u_0}^{u_1}\nonumber\\
&&-(1/2)\int _{u_0}^{u_1} {\dot Q}_{ab}^{\rm el}
(\Psi _4 m^a-r(\eth\Psi _4) l^a)\nonumber\\
 &&\times (l_dm^b-m_dl^b)\, du
+\,\text{conjugate}\, .
\end{eqnarray}
In this form, there are two sorts of problematic terms, each of which will vanish for suitable local restrictions on the matter.  The first are the boundary terms, which involve the spin-coefficients explicitly; these can be eliminated if we consider transitions between $Q_{ab}^{\rm el}=\const$ states.  The other problematic terms are those proportional to $\eth\Psi _4$; those will vanish if we consider quadrupoles which are polarized purely transversely to the waves.  We thus find:

{\em For a gravitational wave encountering a quadrupole of electric type, whose polarization changes only purely transversely to the wave, and making a transition between  two $Q_{ab}^{\rm el}=\const$ states, the energy--momentum exchange is determined by data in the vicinity of the quadrupole, is parallel or anti-parallel to the outgoing direction, and is}
\begin{eqnarray}
\Delta {\mathcal P}_d&=&
-(1/2)\int _{u_0}^{u_1} {\dot Q}_{ab}^{\rm el}
\Psi _4 m^a
 l_dm^b\, du
+\,\text{conjugate}\, .\nonumber\\
&=&-\int _{u_0}^{u_1} {\dot Q}_{ab}^{\rm el}
  C^{apbq}t_pt_q l_d\, du\, ,
\end{eqnarray}
%
{\em where $C_{abcd}$ is the Weyl tensor}.  The qualifications at the beginning of this paragraph give, by contrast, some sense of the degree to which nonlocal considerations can effect the energy--momentum transfer generally.

\subsection{Memory effects}

Even if a system emits gravitational radiation only for a finite interval of retarded time, there may be a net change in its Bondi shear, which is a sort of memory effect.\footnote{The sort of memory figuring here goes back to Bondi \cite{BVM} and is also at the root of Christodoulou's work \cite{Christodoulou1991}.  Indeed, Bondi looked at this in connection with energy--momentum exchange.}  Since the Bondi news $N=-\partial _u{\overline\sigma}_{\rm B}$, the difference in Bondi shear, as a function of angle, will contribute to a holonomy between the regimes before and after the emission of radiation, and thus will have consequences for energy--momentum exchange.

The simplest examples of this occur when mass is ejected from a system.  If a mass $M$ is ejected with four-velocity ${\dot\gamma}^a$, then the resulting change in Bondi shear will be $\Delta\sigma _{\rm B}=2M ({\dot\gamma}\cdot m)^2/{\dot\gamma\cdot l}$ 
in linearized gravity.  For simplicity we consider the effects of this on non-relativistic matter in the radiation zone,
so the dominant contribution to the energy--momentum exchange (\ref{emomex}) comes from $T_{ab}l^al^b$ --- for non-relativistic matter, this is the energy density $\rho$.  Then the exchange will be
\begin{eqnarray}
\Delta {\mathcal P}_d&=&\int T_{ab}l^al^b (\kappa ' m_d
  +\, \text{conjugate} )\, d\tau\nonumber\\
&=&M\int \rho {\dot\gamma}\cdot {\overline m}
\left(\frac{2{\dot\gamma}\cdot{\hat r}{\dot\gamma}\cdot l
 +{\dot\gamma}\cdot\overline{m} \dot\gamma \cdot m}{(\dot\gamma \cdot l)^2} \right)\nonumber\\
&&\times
 m_d r^{-1}\, d^{(3)}{\rm vol}
+\,\text{conjugate}\nonumber\\  
&=&M\int\rho \left( -2 \dot\gamma \cdot{\hat r} (\dot\gamma \cdot l)^{-1}
  +(1/2)\dot\gamma ^p \dot\gamma ^q \Pi _{pq}
  (\dot\gamma \cdot l)^{-2}\right)
\nonumber\\
&&\times  \dot\gamma ^r\Pi _{rd} r^{-1}\, d^{(3)}{\rm vol}
\end{eqnarray} 
Suppose for instance that $\dot\gamma ^a=t^a\cosh\xi +{\hat z}^a\sinh\xi$, and that $\rho$ represents a localized mass $\mu$.  Then 
\begin{eqnarray}\label{DPJ}
\Delta {\mathcal P}^d &=&-\left[\frac{ (3\cos ^2\theta+1)\sinh\xi -4\cos\theta \cosh\xi}{(\cosh\xi -\cos\theta\sinh\xi)^2 } \right]
\nonumber\\
&&\times\frac{M\mu\sin\theta\sinh ^2\xi}{2r}\cdot
 \frac{1}{r}\frac{\partial}{\partial\theta}\, .
\end{eqnarray}
In the ultrarelativistic case, this goes over to
\begin{eqnarray}\label{eltim}
\Delta {\mathcal P}^d 
&=&
  \frac{3\cos\theta -1}{1-\cos\theta}\cdot
    \frac{M\mu\sin\theta e^\xi}{4r}\cdot\frac{1}{r}\frac{\partial}{\partial\theta}\\
    &&
    \qquad\text{for}\quad \theta\gtrsim e^{-\xi}\nonumber\\
 &=& \theta e^{3\xi}
  \cdot
    \frac{M\mu }{2r}\cdot\frac{1}{r}\frac{\partial}{\partial\theta}\\
    &&\qquad\text{for}\quad
\theta \ll e^{-\xi}\, .\nonumber
\end{eqnarray}

Because of its physical interest, let us look at the ultrarelativistic case.  The exchange is everywhere spatial and directed along a meridian.  At colatitudes $\theta <\cos ^{-1}(1/3)=71\deg$ the wave transfers energy--momentum to matter in a direction of decreasing $\theta$, whereas outside this cone the transfer is to increasing $\theta$.  
This will also be the sense of whether the transfer contributes along the direction of the jet or oppositely.  The transfer to the matter will be towards the axis for $\cos ^{-1}(1/3)<\theta <\pi /2$ and away from the axis elsewhere.

Remarkably, a direct calculation from eq. (\ref{DPJ}) shows that for a spherical distribution of matter the net energy--momentum exchange vanishes.  However, this result depends on the cancellation of potentially significant terms.  (For instance, near $\theta =0$ the contributions to the exchange are large in magnitude, but are directed symmetrically in azimuth.)  
From eq. (\ref{eltim}) the fractional energy--momentum exchange will scale like a weighted average over angles of integrals
$(G/c^2)\int\rho rdr$ along the outward null geodesics.
We saw in Section IIIC that such integrals like this could in reasonable astrophysical circumstances be large.
Of course, the present, radiation-dominated, approximation is only valid when these integrals are small. 

We are thus left with the possibility that relativistic jets from sources surrounded by sufficiently inhomogeneous distributions of non-relativistic matter might suffer, within the realm where the analysis here is valid, energy--momentum exchanges which are small (but not very small) fractions of unity.  We cannot say what happens when still more matter is present, but one would very much like to know.  It is certainly possible that grosser effects may occur.

\section{Discussion}

The main idea underlying this paper is that space--times with gravitational radiation are not asymptotically flat in the sense usually required for giving a consistent accounting of energy--momentum, but that this problem is resolved by extending the Bondi--Sachs construction of energy--momentum to the radiation zone.  (This applies even at the linearized level.)  This procedure is strongly non-local, depending on the physics of the gravitational field in all asymptotic directions around the source.  The result is a well-defined way of measuring the contributions of local distributions of matter in the zone to the system's total energy--momentum.  Tracking these gives measures of the exchange of contributions to the total energy--momentum from the matter and gravitational radiation, that is, bulk exchanges, in contrast to the tidal ones usually considered.

This approach is not derived from first principles, but is plausible as long as holonomies in the radiation zone are well-modeled by the leading terms from the Bondi--Sachs analysis.  Since that analysis is concerned with space--times with near-vacuum radiation zones, all of the effects discovered here are relatively small ones in cases where the approach can be conservatively regarded as reliable.  The question of just what happens when we go beyond this ``radiation-dominated'' regime, and in particular, how much larger the energy--momentum exchanges could be, is very much of interest, but will require other techniques, or at least other justifications.

The scattering of test particles was discussed first.  Perhaps surprisingly, although we consider here {\em outgoing} waves, the scattering appeared to be most naturally interpreted in terms of the {\em incoming} congruence associated with the Bondi frame.  We found a shear--shear coupling, between what would be the particle's contribution to the outgoing shear in the linearized limit, and the shear of the incoming congruence, as well as other terms.  Roughly speaking, this is because (for outgoing radiation) the incoming Bondi--Sachs congruence codes the changes in temporal evolution necessary to maintain the Bondi gauge in the presence of outgoing radiation.  The scattering of the particle is due to the adjustments in gauge required by the Bondi--Sachs framework.

The case of test particles scattered by quadrupole radiation in linearized theory was completely worked out.
The details of these results were complex, reflecting partly the freedom in the source's time-dependence, but also the ways the its polarization could couple to the orbital elements of the particles.  Because of this time-dependence, there is no simple universal formula for the fall-off of the scattering with the particles' impact parameters.  For monochromatic sources we found a scaling $\sim (\omega b)^{-2}$ (with $\omega$ the angular frequency and $b$ the impact parameter).  However, for sources slowly varying in the past larger effects were possible.  (In the future, for relativistic particles, retardation provides a cut-off.)

In general, the energy--momentum exchanges will tend to average out if the waves' periods are shorter than the dynamical timescales associated with the matter (intrinsic time-dependences, as well as transit times across the source's sphere of directions).  So I considered several cases in which this averaging could be, a least partially, defeated.

The most straightforward of these was a resonance between the waves and intrinsically time-dependent matter; the latter was modeled by small quadrupoles in the radiation zone.  We did indeed find possibilities for secular energy--momentum exchange.  The effects depended differently on the polarizations than one would find from a na\"ive application of linear theory; this was because the na\"ive picture does not maintain the Bondi--Sachs gauge to the required accuracy.

An issue closely related to resonance is the sense in which gravitational waves' energy--momentum can be effectively directed, absorbed or reflected; I referred to this as {\em dynamic activity}, analogous to optical activity for electromagnetic waves.  (Could one have mirrors, or refractors, of gravitational energy--momentum?  Of course, the work here is limited to fractionally small effects, so it could not justify any very efficient reflection or refraction.)
The nonlocality of the energy--momentum was a serious impediment to having matter which could be dynamically active.  We did find, however, that for matter consisting of small quadrupoles, some sort of localized controlled response was possible if the vector towards the distant source was known; the quadrupoles had to be purely electric, make transitions between constant-quadrupole states, and their changes in polarization had to be transverse to the waves.  The energy--momentum exchange in this case was directed parallel or antiparallel to the wave-vector.

A third way of defeating the averaging was to consider waves with net changes of Bondi shear.  Such waves can exchange energy--momentum even with very simple forms of matter, such as non-relativistic dust.  We considered a simple model, corresponding to waves from a source due to the emission of a relativistic jet.  (The jet is {\em not} the matter with which energy--momentum will be exchanged.)  We found that for perfectly spherically symmetric distributions of matter, the exchange vanished.  However, that was exceptional.  Typically the contribution to the fractional change in energy--momentum of the waves due to matter along a ray outward from the source went like $\sim G\int \rho r\, dr$ (with $\rho$ the density), and a sort of angular average of these was taken.  Such integrals can, in reasonable astrophysical circumstances, become substantial.  This meant that, for inhomogeneous matter around a source due to a jet, the exchange effects can become at least large enough for the present treatment to break down, and the possibility of their being so large as to substantially degrade the waves must be taken seriously.

This leads us to an issue of potentially broad concern: What are the back-reactions on the wave-forms caused by the energy--momentum exchanges?  As pointed out in the introduction, common arguments that such effects will be tiny have involved the implicit assumption that the exchanges are due to tidal effects, but tidal effects would generally be only small fractions of bulk exchanges.  One really needs to revisit the question of the transparency of matter to gravitational waves.
In this connection, one should bear in mind that ultimately gravitational-wave astronomy is expected to require better than per-cent-level accuracy in at least some of the degrees of freedom \cite{LOB2008}.

As emphasized earlier, the present techniques cannot settle this question, for two reasons:  first, they speak to energy--momentum, and not wave-forms; second, they are known to be reliable only in radiation-dominated regimes.  Those regimes are too restricted to count as full realistic models:  a real system might not have the requisite clean geometry, and, even if it does, the waves, as they move outwards,
must eventually weaken to the point that radiation-dominance fails.  However, this paper's analysis can help point to astrophysical situations to investigate by other means, either analytical or numerical.

It does seem very possible that the averaging-out discussed above will mean that in many cases the net modification of the waves will indeed be very small.  Resonance effects could provide secular energy--momentum exchanges and so presumably larger effects, but these require tuning and so are presumably rare.
We also found that for memory effects involving net changes in Bondi shear, where the averaging does not apply, there seemed to be no reason to rule out more substantial back-reaction effects.  
These points could affect the observability of waves from astrophysical jets; compare \cite{BP2013} and references therein.


%

\end{document}